\documentclass[10pt]{article}

\usepackage{amsmath}
\usepackage{amsthm}
\usepackage{amssymb}
\usepackage{graphicx}
\usepackage[hidelinks]{hyperref}
\usepackage{cite}

\usepackage[labelfont=bf,labelsep=period,justification=raggedright]{caption}

\begin{document}

\begin{flushleft}
{\Large
\textbf{TrAp: a Tree Approach for Fingerprinting Subclonal Tumor Composition}
}
\\
Francesco Strino$^{1,\#}$, 
Fabio Parisi$^{1,\#}$, 
Mariann Micsinai$^{1}$,
Yuval Kluger$^{1,\ast}$
\\
\bf{1} Department of Pathology, Yale University School of Medicine, New Haven, Connecticut, United States of America
\\
$\ast$ E-mail: yuval.kluger@yale.edu

$\#$ These authors contributed equally to this work. 
\end{flushleft}

\section*{Abstract}
Revealing the clonal composition of a single tumor is essential for identifying cell subpopulations with metastatic potential in primary tumors or with resistance to therapies in metastatic tumors. Sequencing technologies provide an overview of an aggregate of numerous cells, rather than subclonal-specific quantification of aberrations such as single nucleotide variants (SNVs). Computational approaches to de-mix a single collective signal from the mixed cell population of a tumor sample into its individual components are currently not available. 

Herein we propose a framework for deconvolving data from a single genome-wide experiment to infer the composition, abundance and evolutionary paths of the underlying cell subpopulations of a tumor. The method is based on the plausible biological assumption that tumor progression is an evolutionary process where each individual aberration event stems from a unique subclone and is present in all its descendants subclones. We have developed an efficient algorithm (TrAp) for solving this mixture problem. \textit{In silico} analyses show that TrAp correctly deconvolves mixed subpopulations when the number of subpopulations and the measurement errors are moderate.  We demonstrate the applicability of the method using tumor karyotypes and somatic hypermutation datasets. We applied TrAp to SNV frequency profile from \mbox{Exome-Seq} experiment of a renal cell carcinoma tumor sample and compared the mutational profile of the inferred subpopulations to the mutational profiles of twenty single cells of the same tumor.  Despite the large experimental noise, specific co-occurring mutations found in clones inferred by TrAp are also present in some of these single cells. Finally, we deconvolve \mbox{Exome-Seq} data from three distinct metastases from different body compartments of one melanoma patient and exhibit the evolutionary relationships of their subpopulations.

\section*{Introduction}
The mechanisms of cancer evolution and metastatic onset are still largely unknown. The diversity, complexity and evasive nature of tumor biology are central reasons for the seemingly slow progress in the cure of most cancer types, particularly in controlling the ability of tumor populations to spread. 
Tumor populations are dynamic aggregates of constantly evolving subclones, each carrying a variety of genomic aberrations \cite{NowellClonal, GreavesClonal, Anderson:2011vn, JohnCairns06282006, KleinTumorProgression, Gerlinger:2012uq, Campbell02092008, Hanahan2011646, Podlaha:2012bh, ShahCloneBreast, Mullighan28112008, Notta:2011uq, Driessens:CSC1, Chen:CSC2, Schepers:CSC3, Loeb:2011fk}. This genetic heterogeneity is often associated with differences in the biological behavior of different cell subpopulations. Some of these subclones are likely to be the primary instigators of invasion, metastases or relapse following treatment \cite{Mullighan28112008, Fidler26081977, BJH:BJH2841, Feldman2005435, Shah:2009fk}. An effective characterization of the aggressive potential of tumors at early stages has an enormous potential to guide new clinical interventions and translational research \cite{Fidler:2003kl}.

Recently, several efforts have been made to provide a complete genealogical perspective of cancer evolution \cite{Navin:2011tg,Nolan2004,SingleCell25,SingleCell200Cells,NikZainal2012994}. Using fluorescent labeling techniques, or more recently, single cell sequencing, it is technically possible to separate single cells from tumor samples in order to investigate their evolutionary patterns \cite{NevinSingleCell,Navin:2011tg,Nolan2004,SingleCell25,SingleCell200Cells,NikZainal2012994,
EJH:EJH976,HIS:HIS2188,Yamamoto15112003,AttoliniReview}. However, these approaches are limited to either a small number of fluorescent markers \cite{Nolan2004,NavinFlow}, or to a relatively small number of single cells. On one hand, the identification and selection of uncharacterized subclones in high-throughput experiments is beyond the capabilities of current cell-sorting technologies; on the other hand, isolation and profiling of enough single cells in order to obtain a statistically representative sample of a tumor composed of millions of cells has, currently, prohibitive costs. For this reason, genomics profiling of tumors still relies on pooling in order to provide global averaged signals over the subclonal population within a tumor sample \cite{PLX_Ruth,Banerji2012,Matsushita:2012ij,Varela2011}. Computational methods for identifying subclones, quantifying their relative abundance and monitoring their emergence and dynamics could prove extremely useful for the analysis of the heterogeneity of these pooled samples. This problem has been often overlooked due to its mathematical complexity.

We herein present a mathematical approach to de-mix signals from heterogeneous cell populations into their subclonal components and subsequently unveil the underlying dynamic tumor heterogeneity. Our proposed method is related to the problem of blind source separation \cite{ICA,IFA,LisparseRNA, GongBloodDeconv,BiomarkerDeconvolve,Erkkila:2010qf,ISOLATE, shen2010cell}, where both the underlying sources and their relative composition are unknown. In contrast to blind source separation methods we cannot assume that the underlying sources are statistically independent, we have no prior knowledge of the number of sources and we have at our disposal only one mixture of the unknown sources. This mathematical problem has a vast number of solutions and can be addressed only if additional constraints are imposed. Hence, we assume that tumor cell populations develop in a parsimonious evolutionary process. Furthermore, based on empirical observations, we introduce a sparsity constraint that limits the number of subpopulations. Distinctively from the standard problem of phylogeny \cite{BayesianPhylogeny,Doyon01092011, MixtureTree,MaxPars,TumorMLE,desper1999inferring,desper2000distance,hjelm2006new,radmacher2001graph,
sprouffske2011accurate,Greenman01022012,clement2000tcs}, where each species is observed and measured separately, our methodology, which we term Tree Approach to Clonality (TrAp), is specifically designed to deconvolve a single aggregate signal into its different subclonal components. TrAp incorporates biologically motivated constraints which allow us to infer: 1) the composition of the subclones in a single aggregate sample 2) the abundance of each subclone and 3) the evolutionary path of the subclones.

The paper is organized as follows: we first define the \textbf{subclonal deconvolution problem} and present an efficient algorithm for finding all its solutions. Using \textit{in silico} simulated data we verify that the algorithm is able to correctly deconvolve mixed subclonal populations and that the method is robust to realistic measurement errors. Further, the solution is often unique when the number of populated subclones is moderate. In addition, we also show that TrAp can correctly deconvolve random mixtures of karyotypes of several cells from the same tumor biopsy or from mixture of sequences generated in a study involving somatic hypermutations in B cells. We subsequently compare the mutation profiles of twenty Exome-Seq single cell experiments to those inferred using an aggregate signal generated by exome sequencing from the same renal cell carcinoma tumor. Finally, we apply TrAp to Exome-Seq data from three metastases from three distinct body compartments and compare their subclonal compositions and evolutionary histories.

\section*{Results}
The results are divided in four parts. In the first part we describe our novel TrAp algorithm for subclonal deconvolution of aggregate genomic signals consisting of aberrations' frequencies and we show that the TrAp algorithm always identifies at least one solution; we also introduce an extension to our work that is suited for signals where each locus can be affected by distinct consecutive aberrations (e.g. several consecutive point mutations affecting the same nucleotide). In the second part, we simulate noisy aggregate signals constructed by random linear combinations of simulated subclonal aberration profiles. We used these simulated data to show that TrAp can correctly deconvolve mixtures of evolutionarily related subclones even in presence of levels of noise that are typically found in current genomics experiments. In the third part, we generated realistic aggregate signals by mixing subclonal genomic profiles obtained from cytogenetic analyses using random coefficients. We generate these data separately for each patient and show that, for nearly all aggregate samples, TrAp recovered the subclonal components. Similarly, we generated realistic aggregate signals from somatic hypermutated (SHM) regions from B cells. As we show, somatic hypermutation is a particularly suitable system for the framework of our TrAp algorithm, which successfully recovered all components from the aggregate signals. In the fourth part, we apply our approach to exome-sequencing experiments. We present an analysis of recent single-cell exome sequencing from a renal cell carcinoma study where, besides a collection of twenty single cells, an aggregate has also been measured. Despite the reported difference between the aggregate and mean aberration profile of the single cell experiments, TrAp could still identify subclones with co-occurring aberrations consistent with co-occurring aberrations found in direct single cell measurements. Finally, we analyze three metastases from separate body compartments of a melanoma patient and compare their inferred evolutionary patterns in a genomic region surrounding the \textit{DCC} gene. 

\subsection*{The TrAp algorithm}
\subsubsection*{The subclonal deconvolution problem}
We consider a population of cells where each cell can be described by a binary vector, which we call \textbf{genotype}. Each element of the genotype vector has an aberration state modeled as a binary variable, e.g. the presence/absence of a mutated nucleotide in a specific genomic position or the presence/absence of a specific copy number variation event in a specific locus. For each cell, the $i$-th element of the genotype vector is $1$ if the $i$-th aberration is present in the cell and $0$ if the aberration is absent. A \textbf{subclone} is a collection of all cells that have identical genome-wide aberration profile. A subclone is \textbf{populated} in the sample if the fraction of cells sharing the subclone's genome is greater than zero and can be detected by the experiment. 

\begin{figure}[h]
\begin{center}
\includegraphics[width=4in]{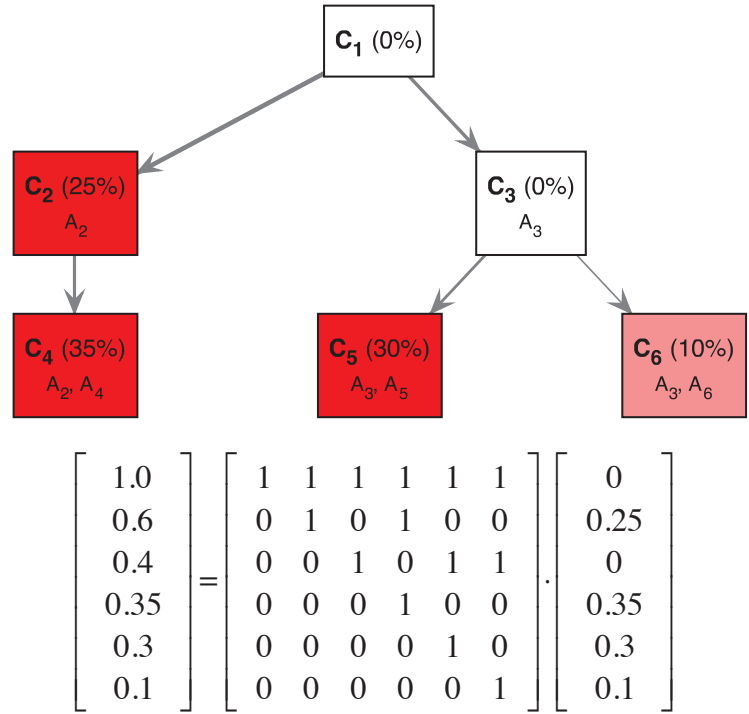}
\end{center}
\caption{
{\bf A schema of deconvolution of the mixed signal of four subclones.} In this example, the aggregate signal frequency vector $\mathbf{y}$ on the left hand side of the matrix-vector equation represents the frequency of five aberrations in the aggregate sample. To allow the heterogeneous mixture of subclones to include normal cells we introduce a dummy aberration that is present in any cell. The frequency of the dummy aberration $y_1$ is equal to one. The frequencies of the actual five aberrations $A_2$, $A_3$, $A_4$, $A_5$, and $A_6$ encoded in the remaining elements of the vector $Y$ are given by $y_2 = 0.6$, $y_3 = 0.4$, $y_4 = 0.35$, $y_5 = 0.3$ and $y_6 = 0.1$, respectively. In this example, the optimal TrAp solution is unique and has four populated subclones: $C_2$ with aberrations $\{A_2\}$, $C_4$ with aberrations $\{A_2, A_4\}$,  $C_5$ with aberrations $\{A_3, A_5\}$ and $C_6$ with aberrations $\{A_3, A_6\}$. The optimal solution is shown both as an evolutionary tree (top) and in matrix form according to Equation~\eqref{EqModel} (bottom), where the tree topology is encoded in the binary matrix and the relative composition of the subclones is represented in the column vector.}
\label{FigTree}
\end{figure}

We define the \textbf{subclonal deconvolution problem} as the task of de-mixing an aggregate measurement into a linear combination of (unknown) subclonal genotypes. The only information that is required as input is the \textbf{aggregate frequency vector} $\mathbf{y}$, whose elements $y_i$ correspond to the frequency of the $i$-th aberration in the sample cell population. For efficiency, we remove from the genome all genotype entries $k$ that are homogenous within the population (i.e. $y_k=0$ or $y_k=1$), as they do not need to be deconvolved. Next, to ensure  that the aberration-free non cancerous cells (wildtype) are included in the solution of the problem, we add one dummy aberration to all the normal and cancerous cells in the sample. By construction, the aggregate frequency of this dummy aberration $y_1$ is equal to $1$. Finally, without loss of generality, we sort the aggregate frequency vector $\mathbf{y}$ in descending order such that $1 = y_1 > y_2 \ge \ldots \ge y_{N} > 0$, where $N$ is equal to the number of aberration events considered including the dummy aberration $y_1$. The subclonal deconvolution problem can be written in matrix notation as

\begin{equation}\label{EqModel}
\mathbf{y}=\mathbf{C} \mathbf{x},
\end{equation}
where $\mathbf{C}$ is a matrix of size $N \times M$ whose elements $c_{i j}$ are $1$ if aberration $i$ is present in subclone $C_j$ and $0$ otherwise; $N$ is the size of the vector $\mathbf{y}$; $M$ is the number of subclones; and $\mathbf{x}$ is a vector of size $M$ where each element $x_j$ corresponds to the frequency of subclone $C_j$ in the sample. We note that, without introducing the wildtype aberration, the wildtype subclone would correspond to a vector of zeros and we would not be able to infer the frequency of the wildtype component
using the linear model of Equation~\eqref{EqModel}. Furthermore, since the dummy aberration is present in the wildtype and all other subclones, it follows that (i)$\forall j,\, c_{1 j} =1$ and (ii) $\sum_j x_j = 1$. We note that since $M$, $\mathbf{C}$ and $\mathbf{x}$ are all unknown in this problem there is an intractable number of possible solutions. As previously discussed \cite{Parisi:2011fk}, for $M>2$ the system is underdetermined and the aggregate signal cannot be uniquely deconvolved by solving the linear system and it is not even possible to find parsimonious unique solutions using sparse reconstruction methods. However, by introducing additional biologically motivated constraints to the model, we can dramatically reduce the number of possible solutions, such that the problem may have a tractable number of optimal solutions, ideally only one. We therefore seek the family of solutions (\textbf{TrAp-solution}s) that sequentially satisfy the following constraints:
\begin{enumerate}
\item \textbf{Evolutionarity}. The subclones are generated in an evolutionary process starting from a subclone with no aberrations. Every subclone is generated from an existing subclone by adding to it a single new aberration event.
\item \textbf{Parsimony}. The number of subclones $M$ that are generated during the evolution process is minimal. 
\item \textbf{Sparsity}. The number of populated subclones $P$ (i.e. the number of subclones $j$ for which $x_j > 0$) is minimal.
\item \textbf{Shallowness}. The depth of the evolutionary tree (i.e. the number of generations) $D=\max_j\left( \sum_i c_{ij} \right)$ is minimal. 
\end{enumerate}
A schema of a TrAp solution is shown in Figure~\ref{FigTree}.

The evolutionarity constraint is used in many biological systems, in particular when studying development of cell populations \cite{BayesianPhylogeny,Doyon01092011, MixtureTree,MaxPars,TumorMLE,desper1999inferring,desper2000distance,hjelm2006new,radmacher2001graph,
sprouffske2011accurate,Greenman01022012}. The parsimony constraint is typically satisfied because the expected probability of a specific aberration event in a nucleotide or a locus is low and it is therefore unlikely that an event occurs more than once and independently in distant subclones. This constraint is the main criterion used to determine the optimality of maximum parsimony methods commonly used in phylogeny \cite{farris1966estimation,kluge1969quantitative,MaxPars,clement2000tcs}. The parsimony constraint dramatically reduces the number of possible solutions of Equation~\eqref{EqModel} because it limits the number of subclones $M$. The sparsity constraint is justified by the fact that some subclones may die out or may be too rare to be detected. Also, it has been shown in several studies that few subclones acquire an evolutionary advantage and outgrow the other subclones \cite{YaosubcloneSize,Malaise1973305,KleinTumorProgression,Driessens:CSC1}, thus reducing the number of populated subclones. The shallowness constraint is justified as excessive genomic instability may not be viable, thus evolutionary trees tend to be shallow and wide rather than deep and narrow.

\subsubsection*{Properties of TrAp solutions}

In this subsection we show that in the subclonal deconvolution problem the evolutionarity and parsimony constraint can always be satisfied by a na\"{i}ve model in which each aberration event occurs exactly once during evolution (i.e. $M=N$). We call any solution with $M=N$ an \textbf{\textit{N}-solution} and its evolutionary tree an \textbf{$N$-tree}. 

The $N$-solution optimally satisfies the evolutionary and parsimony constraints. Since all detected aberrations need to occur at least once in the evolutionary tree, the number of subclones must be greater than or equal to the the number of aberration events considered, i.e. $M \ge N$. We note that it is always possible to construct a valid solution for Equation~\eqref{EqModel} of exactly $M=N$ subclones for every aggregate frequency vector $\textbf{y}$. Specifically, for a cascade-like evolutionary process with no branching, where the wildtype subclone $C_1$ is the root of the tree and every other subclone $C_i$ is a direct descendant of the subclone $C_{i-1}$, the solution of Equation~\eqref{EqModel} is given by:
\begin{equation}\label{EqCascade}
\left[\begin{array}{c}
1\\
y_{2}\\
y_{3}\\
\vdots\\
y_{N}
\end{array}\right]=\left[\begin{array}{ccccc}
1 & 1 & 1 & \cdots & 1\\
0 & 1 & 1 & \cdots & 1\\
0 & 0 & 1 & \cdots & 1\\
\vdots & \vdots & \vdots & \ddots & \vdots \\
0 & 0 & 0 &\cdots & 1
\end{array}\right]\left[\begin{array}{c}
1-y_{2}\\
y_{2}-y_{3}\\
y_{3}-y_{4}\\
\vdots\\
y_{N}
\end{array}\right] 
\end{equation}
While this cascade-like solution satisfies both evolutionarity and parsimony constraints, this solution is not necessarily optimal with respect to the sparsity constraint and is the least optimal in terms of the shallowness constraint. Furthermore, the existence of this solution guarantees that there is always at least one solution to the subclonal deconvolution problem. 

Since we imposed that the four constraints to the subclonal deconvolution problem must be satisfied sequentially, any solution with $M>N$ subclones will always be less optimal than the solution given in Equation~\eqref{EqCascade}, regardless of its sparseness and shallowness. We can thus limit the search space of the TrAp algorithm to $N$-solutions. For this subset of solutions, the vector $\mathbf{x}$ is of size $N$ and $\mathbf{C}$ is a square matrix of size $N \times N$. Importantly, the index of each subclone $C_i$ indicates the subclone for which the $i$-th aberration occurs for the first time. Hence, for any TrAp solution there is a one-to-one correspondence between the $i$-th aberration and the subclone $C_i$ whose index indicates that it evolved from its parent subclone by acquiring the $i$-th aberration. Moreover, each aberration event of an $N$-tree occurs exactly once and cannot be reverted during evolution. Each aberration $i$ is thus present only in subclone $C_i$ and its subclonal descendants:
\begin{equation}\label{Eqxialpha} 
	y_i  = \sum_{j=1}^{N}  c_{ij} x_j = x_i + \sum_{j=1}^{N}  \alpha_{i j} x_j, 
\end{equation}
where $\alpha_{ij}$ is the \textbf{ancestor indicator variable} that is equal to $1$ if subclone $C_i$ is an ancestor of subclone $C_j$ and $0$ otherwise. We note that $\forall j>1, \alpha_{1 j} = 1$, which means that the wildtype clone $C_1$ is always the root of the evolutionary tree, as required by the evolutionarity constraint. 

In Equation~\eqref{Eqxialpha}, we expressed the aggregate frequency $y_i$ as the sum of the frequencies of all subclones descending from $C_i$. We now express the aggregate frequency $y_i$ as a function of the direct descendants of $C_i$. We define the \textbf{parent indicator variable} $\phi_{ij}$, which is $1$ if $C_i$ is the parent of $C_j$ (i.e. if subclone $C_j$ is the result of a single aberration event in subclone $C_i$) and $0$ otherwise. Finally, using the parent indicator variables we express $y_i$ in terms of the aggregate frequencies of the direct descendants of $C_i$
\begin{equation}\label{Eqxiphid}
	y_i  = x_i + \sum_{j=1}^{N}  \alpha_{i j} x_j = x_i + \sum_{j=1}^{N} \phi_{i j} \left( x_j + \sum_{k=1}^{N} \alpha_{j k} x_k \right)  = x_i + \sum_{j=1}^{N} \phi_{i j} y_j,
\end{equation}

Equation~\eqref{Eqxiphid} can be rearranged to express the vector $\mathbf{x}$ in terms of $\mathbf{y}$ and the parent indicator matrix $\mathbf{\Phi}$:
\begin{equation}\label{Eqxiphi}
	\mathbf{x} = \mathbf{y}-\mathbf{\Phi} \mathbf{y},
\end{equation}
where $\mathbf{\Phi}$ is the $N\times N$ matrix whose elements are given by the parent indicator variables $\phi_{i,j}$. Because of the evolutionary constraint, the matrix $\mathbf{\Phi}$ has only $N-1$ nonzero elements, reflecting the fact that each subclone except the wildtype has exactly one parent subclone, i.e. $\sum_{i=1}^{N}\phi_{i1}=0$ and $\forall j>1,\, \sum_{i=1}^{N}\phi_{ij}=1$. Furthermore, an important corollary of Equation~\eqref{Eqxiphid} is that the subclone $C_i$ is not populated if and only if
\begin{equation}\label{Eqxiphi0}
	y_i - \sum_{j=1}^{N} \phi_{i j} y_j =0.
\end{equation}
In other words, the clone $C_i$ is not populated when the aggregate frequency $y_i$ of aberration $i$ is equal to the sum of the aggregate frequencies of all the direct descendants of the subclone $C_i$. Therefore, the number of non-populated subclones of the $N$-tree encoded by $\mathbf{\Phi}$ is given by the number of aberrations $i$ that satisfy Equation~\eqref{Eqxiphi0}. 

Finally, we summarize the relationships between $\mathbf{C}$ and the indicator variables $\alpha$ and $\phi$. Using Equation~\eqref{Eqxialpha}, we can express the subclonal deconvolution problem as $\mathbf{y}= \mathbf{C}\mathbf{x} = (\mathbf{I} + \mathbf{A}) \mathbf{x}$, where $\mathbf{I}$ is the $N\times N$ identity matrix and $\mathbf{A}$ is the $N\times N$ matrix of whose elements are given by the ancestor indicator variable $\alpha_{ij}$. Furthermore, Equation~\eqref{Eqxiphi} allows us to write the subclonal deconvolution problem as $\mathbf{x} = (\mathbf{I} - \mathbf{\Phi}) \mathbf{y}$. We can therefore express the relationships between the matrices $\mathbf{C}$, $\mathbf{A}$ and $\mathbf{\Phi}$ as:
\begin{equation}\label{EqVecMatTree}
	\mathbf{C} = (\mathbf{I} + \mathbf{A}) = (\mathbf{I} - \mathbf{\Phi})^{-1}.
\end{equation}

\subsubsection*{A brute-force algorithm for solving the subclonal deconvolution problem}

We now observe that the relationship between two subclones $C_i$ and $C_j$ such that $i<j$ (which also implies $y_i \ge y_j$ as the vector $\mathbf{y}$ is sorted), must be one of the following: i) $C_i$ is an ancestor of $C_j$, i.e. $\alpha_{i j}=1$, $\alpha_{j i}=0$ and $y_i \ge y_j$; or ii) $C_i$ and $C_j$ are on separate branches, i.e. $\alpha_{i j}=0$, $\alpha_{j i}=0$ and $y_i + y_j \le 1$. This property implies that all evolutionary trees can be generated iteratively by starting with the wildtype clone $C_1$ and adding the clone $C_i$ at step $i$ to all trees generated at step $i-1$. In detail, for any tree that can be generated using subclones $C_1, \ldots, C_{i-1}$ we generate a new tree by adding the subclone $C_i$ as direct descendant of subclone $C_j$ for all $j<i$ for which the resulting $x_j$ (calculated using Equation~\eqref{Eqxiphi}) remains nonnegative after adding $C_i$. 

\begin{figure}[!ht]
\begin{center}
\includegraphics[width=\linewidth]{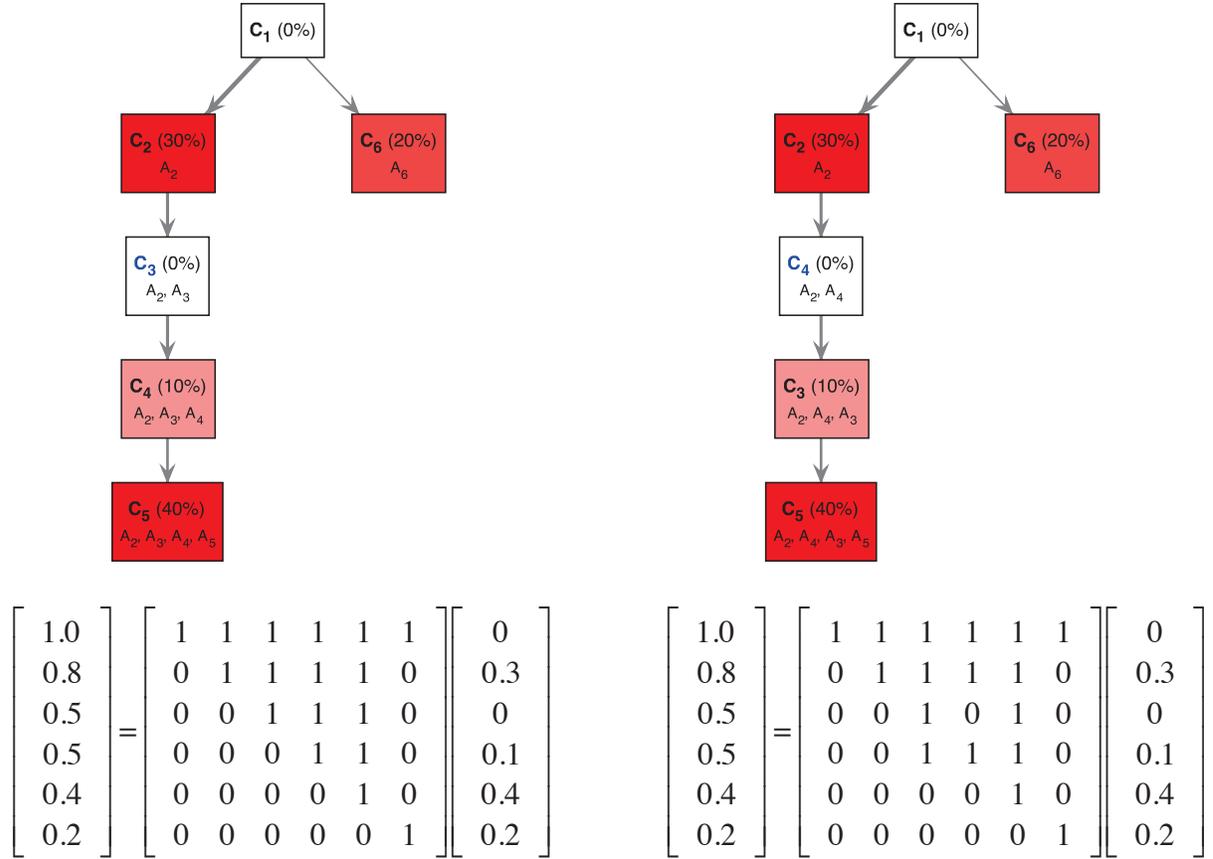}
\end{center}
\caption{
{\bf Deconvolution of a mixture where two aggregate signal frequencies are equal.} In this example, five aberrations ($A_2$, $A_3$, $A_4$, $A_5$ and $A_6$) were measured from an aggregate sample and their frequencies were $y_2 = 0.8$, $y_3 = 0.5$, $y_4=0.5$, $y_5=0.4$ and $y_6=0.2$, respectively. The dummy measurement $y_1 = 1$ was also added to generate the aggregate signal frequency vector $\mathbf{y}=[1, 0.8, 0.5, 0.5, 0.4, 0.2]$. In this example, there are two optimal TrAp solutions (left and right), each shown both as an evolutionary tree (top) and in matrix form according to Equation~\eqref{EqModel} (bottom). Both solutions have  $4$ common populated subclones, namely $C_2$ with aberration $\{A_2\}$, $C_5$ with aberrations $\{A_2, A_3, A_4, A_5\}$, $C_6$ with aberration $\{A_6\}$ and a subclone with aberrations $\{A_2, A_3, A_4\}$. In both cases, the ancestors of the clone with aberrations $\{A_2,A_3,A_4\}$ ($C_3$ of the left tree and $C_4$ of the right tree) are not populated. We remark that these two solutions are practically  indistinguishable and that the TrAp algorithm outputs only the solution where the subclonal indices along each branch are arranged in an increasing order (as shown in the left tree solution). }
\label{FigPerm}
\end{figure}

\begin{figure}[!ht]
\begin{center}
\includegraphics[width=\linewidth]{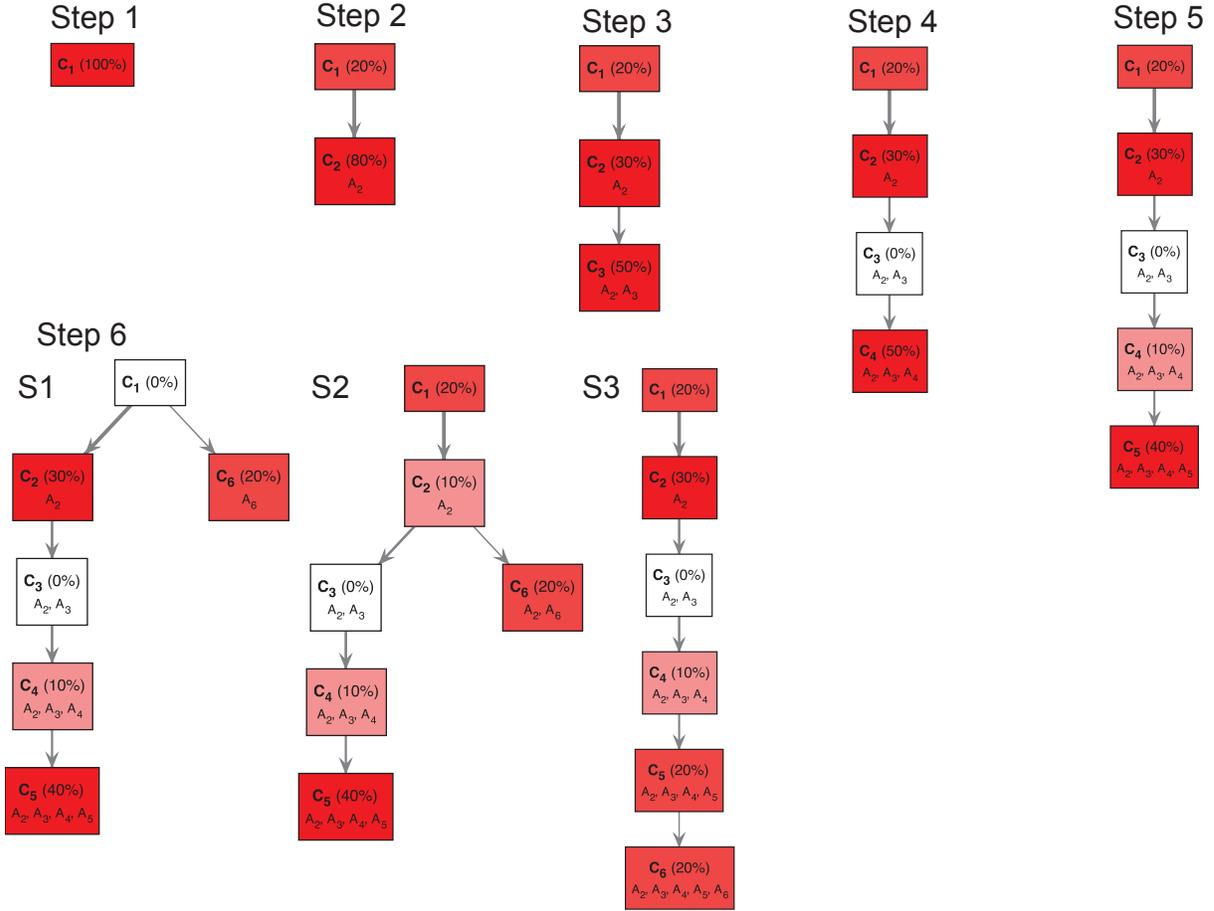}
\end{center}
\caption{
{\bf Brute-force search approach to deconvolve a mixture of four subclones.} In this example, five aberrations ($A_2$, $A_3$, $A_4$, $A_5$ and $A_6$) were measured from an aggregate sample and their frequencies were $y_2 = 0.8$, $y_3 = 0.5$, $y_4=0.5$, $y_5=0.4$ and $y_6=0.2$, respectively. The dummy measurement $y_1 = 1$ was also added to generate the aggregate signal frequency vector $\mathbf{y}=[1, 0.8, 0.5, 0.5, 0.4, 0.2]$. 
In the first step, the wild type clone $C_1$ (representing the wildtype subpopulation) is positioned at the root of the tree. In the second step, a tree reconstruction begins and $C_2$ is added to the only possible ancestor clone $C_1$. In the third step, $C_3$ must be added as direct descendant of $C_2$. $C_3$ cannot be added to the tree on a different branch as a direct descendant of $C_1$ because based on Equation~\eqref{Eqxiphi} this would imply a negative frequency $x_1=y_1-y_2-y_3=-0.3$. Likewise, the subclones $C_4$ and $C_5$ can only be added as direct descendants of $C_3$ and $C_4$, respectively. Finally, $C_6$ can be added as a direct descendant of subclones $C_1$, $C_2$ or $C_5$ generating solutions $S_1$, $S_2$ or $S_3$, respectively. However, $S_1$ is the only TrAp-solution as its number of populated subclones is minimal and corresponds the solution shown in the left side of Figure~\ref{FigPerm}. Solution $S_3$ is the cascade-like solution described in Equation~\eqref{EqCascade}. }
\label{FigBF}
\end{figure}

For completeness, when $y_i = y_j$ the subclones $C_i$ and $C_j$ can be either on separate branches or on the same branch.  If they are on the same branch, then there is an ambiguity regarding the order in which the two aberrations occur (Figure~\ref{FigPerm}). However, in the case that these two subclones are on the same branch, the aberration profile of the ancestor subclone (shown in green in Figure~\ref{FigPerm}) is not informative because this subclone is not populated ($x_a=0$) and aberrations $i$ and $j$ are both present  in the descendant subclone regardless of the order in which they occur. Since these two aberrations cannot be observed separately (i.e. the coefficient $x_a$ associated with the ancestor aberration is zero, whereas the $x_{d}$ associated with both aberrations could be nonzero), we only output the solution for which $C_{\min\{i,j\}}$ is an ancestor of $C_{\max\{i,j\}}$ (left solution in Figure~\ref{FigPerm}). This choice ensures that for every pair of aberrations $i < j$, $C_j$ cannot be an ancestor of $C_i$. A step-by-step example of the TrAp solution obtained using the brute-force algorithm is given in Figure~\ref{FigBF}.

The upper bound on the number of possible evolutionary $N$-trees is thus $(N-1)!$, as every subclone $i$ can only be the direct descendant of $i-1$ subclones. This bound is significantly smaller than the number of all possible trees with $N$ labeled vertices, which is $N^{N-2}$ according to Cailey's formula \cite{caileyformulaORG,CayleyFormula}. Furthermore, we note that the parent indicator matrix $\mathbf{\Phi}$ and $\mathbf{C}$ are upper triangular and that both $\mathbf{C}$ and $\mathbf{I}-\mathbf{\Phi}$ are of rank $N$ and invertible, which guarantees that Equation~\eqref{EqVecMatTree} can always be used to switch between the representation with the parent indicator variable $\mathbf{\Phi}$ (Equation~\eqref{Eqxiphi}) and the representation with the subclone matrix $\mathbf{C}$ (Equation~\eqref{EqModel}). We also note that, given a matrix $\mathbf{\Phi}$ (or $\mathbf{C}=(\mathbf{I}-\mathbf{\Phi})^{-1}$), the vector $\mathbf{x}$ is uniquely determined. In particular, if the $\mathbf{C}$ matrix is known, the vector $\mathbf{x}$ can be efficiently found by solving the linear system $\mathbf{C}\mathbf{x}=\mathbf{y}$ using back-substitution (i.e. by solving Equation~\eqref{Eqxialpha} first for $x_N$, then using $x_N$ to solve for $x_{N-1}$ and repeating through $x_1$). 

\subsubsection*{TrAp: a fast algorithm for solving the subclonal deconvolution problem}

As we have shown, the number of non-populated subclones of a given $N$-tree is equal to the number of aberrations $i$ that satisfy Equation~\eqref{Eqxiphi0}. Moreover, in order to satisfy the sparsity constraint of a solution, we do not need to know the topology of the whole evolutionary tree, but only the subset of rows of the matrix $\mathbf{\Phi}$ that satisfy Equation~\eqref{Eqxiphi0}. We now leverage on this property to efficiently generate sparse solutions to the subclonal deconvolution problem. 

\begin{figure}[!ht]
\begin{center}
\includegraphics[width=\linewidth]{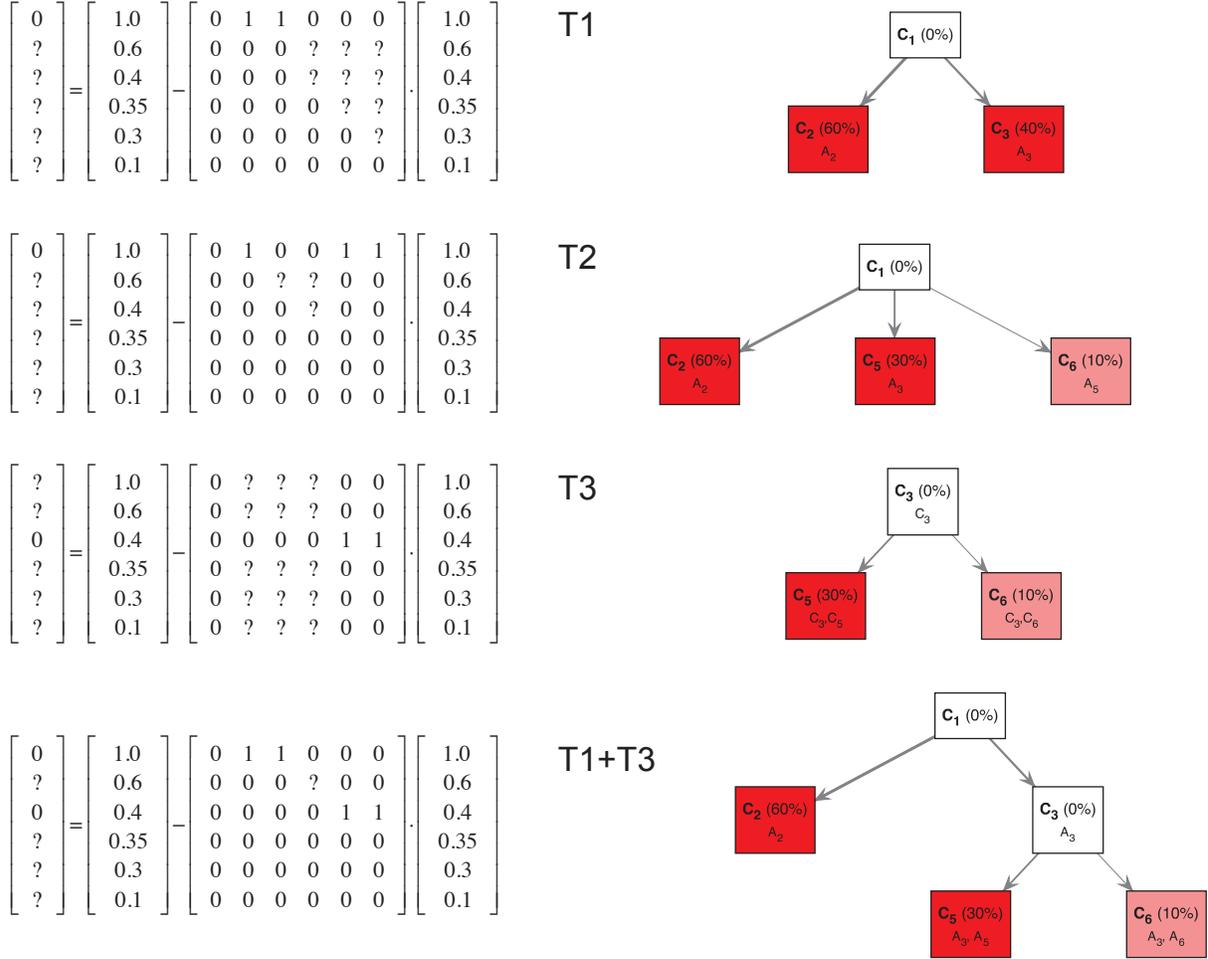}
\end{center}
\caption{
{\bf Identification of first generation trees.} In this example, the aggregate signal frequency vector $\mathbf{y}=[1,0.6,0.4,0.35,0.3,0.1]$ is consistent with three first generation trees $T_1=\{C_1,C_2,C_3\}$, $T_2=\{C_1,C_2,C_5,C_6\}$ and $T_3=\{C_3,C_5,C_6\}$. Each first generation tree is visualized as a matrix equation $X=Y-\Phi Y$ according to Equation~\eqref{Eqxiphi} (left) and as a partial evolutionary tree (right). In the bottom row, the partial tree $PT_1$ given by the union of the partial trees $T_1$ and $T_3$ is shown. Question marks indicate values that are unknown as they are not specified by the first generation tree or by the partial tree.}
\label{FigFGT}
\end{figure}

First, we group each subset of subclones that satisfy Equation~\eqref{Eqxiphi0} into a \textbf{first generation tree} $T_i$, which we define as the subset of subclones $\left\{C^{T_i}_p, C^{T_i}_1, \ldots,  C^{T_i}_{N_i}  \right\}$ such that the subclone $C^{T_i}_p$ is not populated (i.e. $x^{T_i}_p=0$) and that $N_i$ subclones $C^{T_i}_1,\ldots,C^{T_i}_{N_i}$ are its direct descendants. Each first generation tree is represented by a row of the $\mathbf{\Phi}$ matrix. For example, there are three first generation trees for the aggregate signal $Y=\{1, 0.6, 0.4, 0.35, 0.3, 0.1\}$, namely $T_1=\{C_1,C_2,C_3\}$, $T_2=\{C_1,C_2,C_5,C_6\}$ and $T_3=\{C_3,C_5,C_6\}$ (Figure~\ref{FigFGT}). We note that the optimal TrAp solution for this example contains the first generation trees $T_1$ and $T_3$ (Figure~\ref{FigTree}). Furthermore, a $\mathbf{\Phi}$ matrix associated with a first generation tree must follow the evolutionary constraints previously described ($\forall j>1,\, \sum_{i=1}^{N}\phi_{ij}=1$) and thus the first generation tree also gives complete information about the columns of $\mathbf{\Phi}$ corresponding to the direct descendant subclones $C^{T_i}_1, \ldots,  C^{T_i}_{N_i}$. For example, the first generation tree $T_1=\{C_1,C_2,C_3\}$ implies that $\phi_{i,2}=0$ and $\phi_{i,3}=0$ for every $i \ne 1$ (Figure~\ref{FigFGT}).

Next, we define a \textbf{partial tree} as a collection of first generation trees $\{T_1, \ldots, T_h\}$ that can jointly be contained in a full evolutionary tree. Since each first generation tree can be represented by a row of the $\mathbf{\Phi}$ matrix, a partial tree that is comprised of $h$ first generation trees can be represented by $h$ rows of the $\mathbf{\Phi}$ matrix. In the example above, the partial tree that contains the first generation trees $T_1$ and $T_3$ is represented by rows $1$ and $3$ of the matrix $\mathbf{\Phi}$ in the bottom panel of Figure~\ref{FigFGT}. Similarly to first generation trees, the matrix $\mathbf{\Phi}$ associated with a partial tree must follow the evolutionary constraint, which implies that not all combinations of first generation trees give rise to partial trees. In the example above, the partial trees $T_1$ and $T_3$ can be combined to generate the partial tree $PT_1=\{T_1, T_3\}$ (Figure~\ref{FigFGT} bottom), whereas the pairs $\{T_1, T_2\}$ and $\{T_2, T_3\}$ cannot be combined to generate partial trees. Therefore, in the example above the possible partial trees are $PT_1=\{T_1, T_3\}$, the empty partial tree $PT_2=\{\}$ and the partial trees $PT_3=\{T_1\}$, $PT_4=\{T_2\}$ and $PT_5=\{T_3\}$. Moreover, we note that all $N$-trees that contain a partial tree comprising of $h$ first generation trees have $N-h$ populated subclones. This implies that TrAp solutions, which must satisfy the sparsity constraint, must also contain one of the partial trees comprising the maximum number of first generation trees. In the example above, the optimal TrAp solution (Figure~\ref{FigTree}) contains the partial tree $PT_1$, which is the only partial tree comprising two first generation trees.

\begin{figure}[!ht]
\begin{center}
\includegraphics[width=\linewidth]{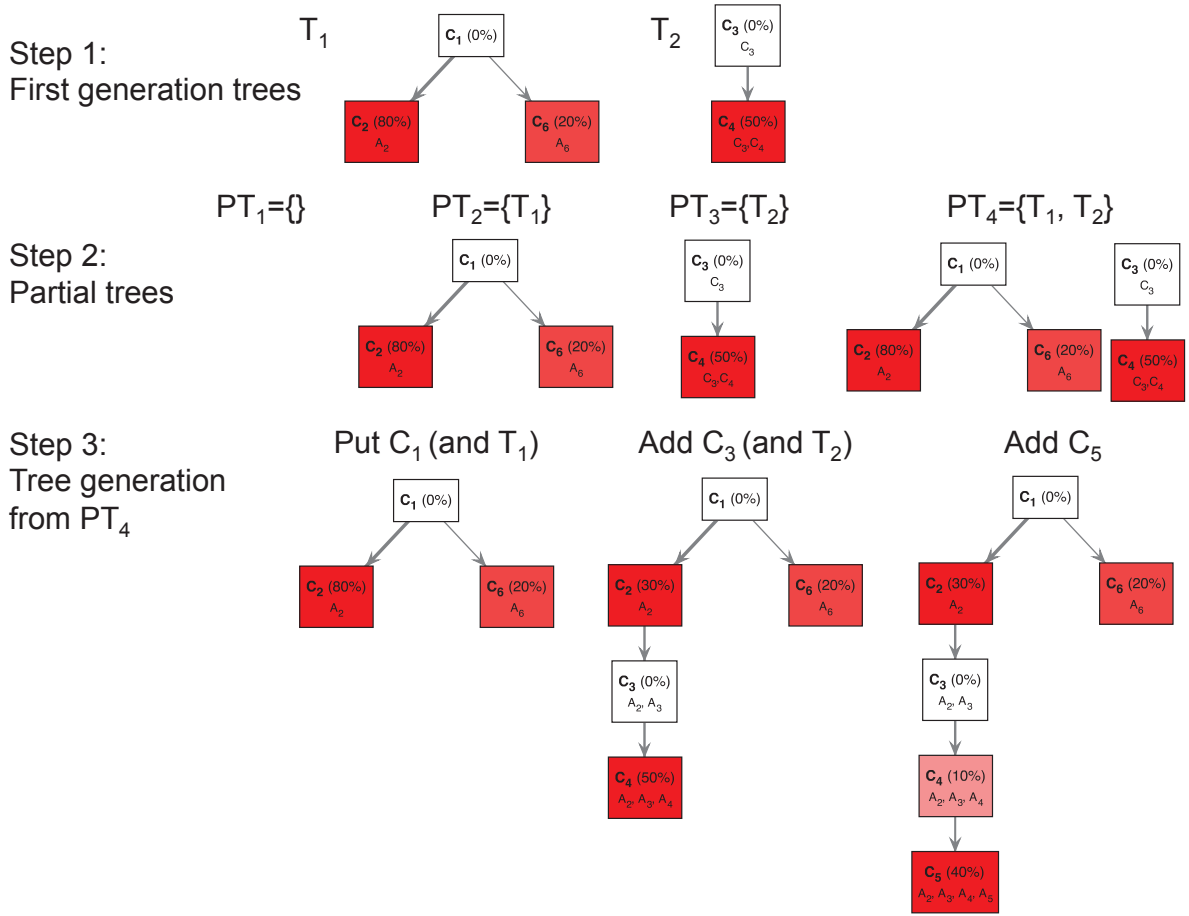}
\end{center}
\caption{
{\bf Illustration of the usage of first generation trees and partial trees for deriving the TrAp solution of a mixture of four subclones.} In this example, five aberrations were measured from an aggregate sample and their frequencies were $y_2 = 0.8$, $y_3 = 0.5$, $y_4=0.5$, $y_5=0.4$ and $y_6=0.2$, respectively. The dummy measurement $y_1 = 1$ was also added to generate the aggregate signal frequency vector $\mathbf{y}=[1, 0.8, 0.5, 0.5, 0.4, 0.2]$. 
In the first step, TrAp identifies all first generation trees, namely $T_1=\{C_1,C_2,C_6\}$ and $T_2=\{C_3,C_4\}$. In the second step, TrAp generates the possible partial trees, namely $PT_1=\{\}$, $PT_2=\{T_1\}$, $PT_3=\{T_2\}$ and $PT_4=\{T_1,T_2\}$, and consequently selects only $PT_4=\{T_1,T_2\}$ as it is the only partial tree which contains a maximum number of first generation trees.
In the third step, TrAp generates evolutionary trees starting from the partial tree $PT_4=\{T_1,T_2\}$. To complete the evolutionary tree starting from $PT_4$, the subclone $C_1$ is positioned as the root of the tree. Since $C_1$ is part of the first generation tree $T_1$, the subclones $C_2$ and $C_6$ are automatically added as a direct descendant of $C_1$. Next, $C_3$ is added as a direct descendant of $C_2$. Since $C_3$ is part of the first generation tree $T_2$, $C_4$ is automatically added as direct descendant of $C_3$. Finally, $C_5$ is added as a direct descendant of $C_4$, generating the optimal TrAp solution to the subclonal deconvolution problem. We remark that the optimal solution generated by the TrAp algorithm is equal to $S1$ in Figure~\ref{FigBF} and to the left solution of Figure~\ref{FigPerm}.}
\label{FigTrAp}
\end{figure}

Since all TrAp solutions contain the maximum number of first generation trees, the TrAp algorithm dramatically reduces the search space by identifying the optimal partial trees and later utilizing them as starting points for rapidly building all the sparsest solutions to the subclonal deconvolution problem. In details, the TrAp algorithm solves the subclonal deconvolution problem in the following steps (Figure~\ref{FigTrAp}):
\begin{enumerate}
\item Identify all first generation trees from the aggregate signal vector $\mathbf{y}$.
\item Combine all first generation trees to generate all partial trees. 
\item Discard partial trees that do not have the minimum number of populated subclones. 
\item Generate all evolutionary trees consistent with the partial trees comprising the maximum number of first generation trees. This step is done iteratively for each partial tree, similarly to the way described for the brute-force algorithm. The only difference is that, when the parent clone $C^{T_i}_p$ of a first generation tree $T_i=\left\{C^{T_i}_p, C^{T_i}_1, \ldots,  C^{T_i}_{N_i}  \right\}$ is added to the tree, the subclones $C^{T_i}_1, \ldots,  C^{T_i}_{N_i}$ are automatically added as direct descendants of $C^{T_i}_p$.
\item Optimize the shallowness constraint by sorting the generated solutions by the depth of the generated tree.
\end{enumerate}

The performance of the TrAp algorithm is equivalent to the brute-force approach when there are no first generation trees (i.e. when all subclones are populated) and becomes superior to a brute-force approach when $P<N$. Furthermore, the algorithm can be modified by imposing additional constraints that need to be satisfied at each step of the iterative tree reconstruction procedure (an example of such modification is presented in the subsection on extension of TrAp to non-binary aberrations). The TrAp algorithm by default returns only the solution(s) that optimize all the constraints. In addition, the user can specify parameters to relax the sparsity and shallowness constraints. 

\subsubsection*{Generalization of TrAp to non-binary aberrations}

In the previous sections we represented the genome of each subclone by a vector of binary values whose entries represent if a genomic position is in a normal state (0) or in an aberrated state (1). In general the number of  states in a given genomic position could be larger than two and hence subclones cannot be represented by vectors of binary values without loss of information. For example, a nucleotide found in the reference genome or in the germline at a specific position may undergo multiple distinct point mutation events into more than one specific nucleotide. In this subsection, we describe an extension of the TrAp algorithm to deal with such cases. 

\begin{figure}[!ht]
\begin{center}
\includegraphics[width=\linewidth]{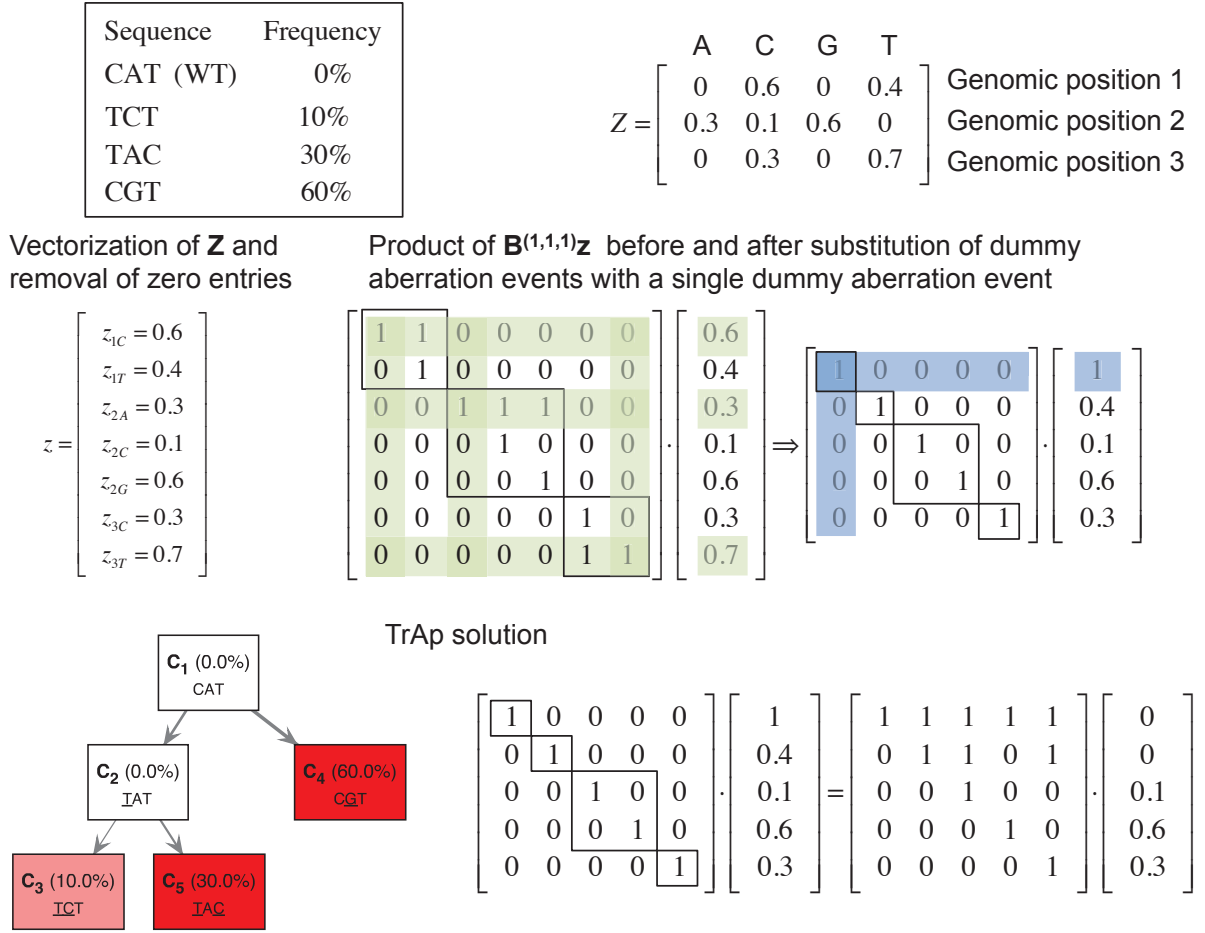}
\end{center}
\caption{
{\bf Application of the TrAp algorithm to deconvolve a mixture of three sequences in presence of poly-allelic mutations.} We analyzed an aggregate sample composed of three subclones with sequences "TCT", "TAC" and "CGT" mixed with frequencies $0.1$, $0.3$ and $0.6$, respectively. In this particular example, the wildtype sequence  "CAT" is absent in the mixture. Nonzero elements of the aggregate frequency matrix $\mathbf{Z}$ (top right) are then concatenated in the $\mathbf{z}$ vector, which consists of the elements $z_{1C}=0.6$, $z_{1T}=0.4$, $z_{2A}=0.3$, $z_{2C}=0.1$, $z_{2G}=0.6$, $z_{3C}=0.3$ and $z_{3T}=0.7$ (middle left). In the center of the middle panel we show the binarization matrix $\mathbf{B^{(1,1,1)}}$ and the matrix-vector product $\mathbf{B^{(1,1,1)}} \mathbf{z}$ associated with it, which are consistent with Equation~\eqref{EqB} and lead to the optimal TrAp solution. Next, rows and columns corresponding to unmutated states (i.e. $1C$, $2A$ and $3T$, shown in green) are substituted with a  dummy aberrated state corresponding to the wildtype (shown in blue). In the bottom row, the TrAp solution is shown both as an evolutionary tree (left) and in matrix form according to Equation~\eqref{EqGeneralModel} (right). }
\label{FigCAT}
\end{figure}

We consider the \textbf{generalized subclonal deconvolution problem} in which the genome consists of $N$ positions each of which can have $S$ different states. We also assume that the genome of the wildtype subclone is known. The only information that we utilize as input is the \textbf{aggregate frequency matrix} $\mathbf{Z}$ whose elements $z_{k,s}$ correspond to the observed frequency of the aberrated state $s$ at position $k$. We note that, by construction, $0 \le z_{k,s} \le 1$ and that $\sum_{s=1}^{S} z_{ks} = 1$. To utilize our framework for solving the subclonal deconvolution problem, we convert the information encoded in $\mathbf{Z}$ as a vector $\mathbf{y}$ whose elements represent frequencies of binary events. We perform this transformation in several steps. First, we vectorize the matrix $\mathbf{Z}$ by concatenating its rows to construct the vector $\mathbf{z}$, which has $K S$ elements. Then, we remove from the vector $\mathbf{z}$ the entries for which $z_{ks}=0$ as they are not informative. As a result, for each position $k$ there are $S_k$ entries, where $S_k$ ranges from $1$ (only the unmutated state is observed) to $S$ (all aberrated states are observed). For illustration of these first steps, we consider a toy example of a genome of length three whose wildtype sequence is "CAT" and we analyze an aggregate sample made of three subclones with sequences "TCT", "TAC" and "CGT" mixed with frequencies $0.1$, $0.3$ and $0.6$, respectively (Figure~\ref{FigCAT}). In this example the $\mathbf{z}$ vector consists of the elements $z_{1C}=0.6$, $z_{1T}=0.4$, $z_{2A}=0.3$, $z_{2C}=0.1$, $z_{2G}=0.6$, $z_{3C}=0.3$ and $z_{3T}=0.7$ (Figure~\ref{FigCAT}).

Next, we wish to design a binarization matrix $\mathbf{B}$ to encode the information contained in $\mathbf{z}$ as a vector $\mathbf{y}=\mathbf{B}\mathbf{z}$ whose elements represent the frequency of binary aberrations and can thus be used as input to the subclonal deconvolution problem for the whole genome (Equation~\eqref{EqModel}). For every position $k$, we assume that each state $s$ ($1\le s\le S_k$) is reached by a sequence of aberration events. We denote by $A_{ks}$ an aberration event to state $s$ at position $k$. In the example above, there are two states at position $1$. The unmutated state $C$ is reached by a dummy aberration $A_{1C}$ (in analogy to the dummy aberration of the wildtype clone in the subclonal deconvolution problem) and the aberrated state $T$ is reached by the sequence of the dummy aberration $A_{1C}$ followed by the aberration $A_{1T}$. Since the dummy aberration $A_{1C}$ is present when we observe states $C$ or $T$ at position $1$, the frequency of the unmutated aberration is $y_{1C}=z_{1C}+z_{1T}=1$. However, the aberration $A_{1T}$ is present only when we observe the state $T$, therefore the frequency of the aberration $A_{1T}$ is $y_{1T}=z_{1T}=0.4$ (Figure~\ref{FigCAT}). Since measurements at different genomic positions do not affect one another, we construct $\mathbf{B}$ as a block-diagonal matrix:
\begin{equation}\label{EqB}
\left[\begin{array}{c}
\mathbf{y_{1}}\\ \hline
\mathbf{y_{2}}\\ \hline
\vdots \\ \hline
\mathbf{y_{K}}
\end{array}\right]=
\left[\begin{array}{cccc}
\cline{1-1}
\multicolumn{1}{|c} {\mathbf{B_1}} &\multicolumn{1}{|c} {0} & \cdots & 0 \\ \cline{1-2}
0  & \multicolumn{1}{|c} {\mathbf{B_2}}  & \multicolumn{1}{|c} {\cdots} & 0\\   \cline{2-3}
\vdots & \vdots & \multicolumn{1}{|c} {\ddots}  & \multicolumn{1}{|c} {\vdots} \\ \cline{3-4}
0 & 0  &\cdots & \multicolumn{1}{|c|} {\mathbf{B_K}} \\ \cline{4-4}
\end{array}\right] 
\left[\begin{array}{c}
\mathbf{z_{1}}\\ \hline
\mathbf{z_{2}}\\ \hline
\vdots \\ \hline
\mathbf{z_{K}}
\end{array}\right],
\end{equation}
where $\mathbf{y_k}=\left[y_{k1}, \ldots, y_{kS_k} \right]$ and $\mathbf{zk}=\left[z_{k1}, \ldots, z_{kS_k} \right]$. Combining Equation~\eqref{EqModel} and Equation~\eqref{EqB} gives
\begin{equation}\label{EqGeneralModel}
\mathbf{B} \mathbf{z} = \mathbf{C} \mathbf{x}.
\end{equation}
It is important to note that for any pair of different states $s_1$ and $s_2$ at position $k$ where $1 \le s_1 \le S_k$ and $1 \le s_2 \le S_k$, the value of $b_{ks_1,ks_2}$ defines the ancestral relationship between the aberration events $A_{ks_1}$ and $A_{ks_2}$. Using the ancestor indicator variable $\alpha$ we can express this relationship as $\alpha_{ks_1,ks_2}=b_{ks_1,ks_2}$. To preserve these ancestral relationships in both sides of Equation~\eqref{EqGeneralModel} (we recall that $\mathbf{C}=\mathbf{I}+\mathbf{A}$), the element $c_{ks_1,ks_2}$ of $\mathbf{C}$ must be equal to the element $b_{ks_1,ks_2}$ of $\mathbf{B}$ for every pair of states $s_1$ and $s_2$ at position $k$, where $1 \le k \le K$, $1 \le s_1 \le S_k$ and $1 \le s_1 \le S_k$.

In order to find the solutions to Equation~\eqref{EqB}, we solve independently each subproblem  $\mathbf{y_k}=\mathbf{B_k}\mathbf{z_k}$ and we require that the $\mathbf{B_k}$ matrix encodes a tree which satisfies the evolutionary and parsimony constraints and whose root is the dummy unmutated aberration at position $k$. The number of solutions for each subproblem is equal to the number of trees with $S_k$ labeled vertices and, as discussed earlier, this number is equal to $S_{k}^{S_k-2}$. We then denote by $\mathbf{B_{k}^{(t_k)}}$ the $t_k$-th solution ($1 \le t_k \le S_k^{S_k-2}$) and by $\mathbf{y_k^{(t_k)}}$ the aggregate frequency vector associated to it. The solutions to the problem for $S_k \le 3$ are:

\begin{description}
  \item[$S_k=1$.] There is only one solution. Since the input consists only of the unmutated state, it follows that $\mathbf{z_{k}}=\left[1\right]$. The solution is $\mathbf{B_k^{(1)}}=[1]$ and the corresponding aggregate frequency vector is $\mathbf{y_{k}^{(1)}}=\left[1\right]$. 

  \item[$S_k=2$.] There is only one solution. Assuming the input is $\mathbf{z_{k}}=\left[z_{k1}, z_{k2} \right]$ and the unmutated state is $k1$, the solution is $\mathbf{B_k^{(1)}}=\left[\begin{array}{cc}1 & 1 \\ 0 & 1  \end{array}\right]$ and the corresponding aggregate frequency vector is $\mathbf{y_{k}^{(1)}}=\left[1, z_{k2} \right]$.

  \item[$S_k=3$.] There are $3$ solutions. Assuming the input is $\mathbf{z_{k}}=\left[z_{k1}, z_{k2}, z_{k_3} \right]$ and unmutated state is $k1$, the solutions are $\mathbf{B_k^{(1)}}=\left[\begin{array}{ccc}1 & 1 & 1 \\ 0 & 1 & 0 \\ 0 & 0 & 1 \end{array}\right]$, $\mathbf{B_k^{(2)}}=\left[\begin{array}{ccc}1 & 1 & 1 \\ 0 & 1 & 1 \\ 0 & 0 & 1 \end{array}\right]$ and $\mathbf{B_k^{(3)}}=\left[\begin{array}{ccc}1 & 1 & 1 \\ 0 & 1 & 0 \\ 0 & 1 & 1 \end{array}\right]$and their corresponding $\mathbf{y_{k}}$ vectors are given by $\mathbf{y_{k}^{(1)}}=\left[1, z_{k2}, z_{k3} \right]$, $\mathbf{y_{k}^{(2)}}=\left[1, z_{k2}+ z_{k3}, z_{k3} \right]$ and $\mathbf{y_{k}^{(3)}}=\left[1, z_{k2}, z_{k2}+z_{k3} \right]$. The values $b_{k2,k3}$ and $b_{k3,k2}$ define the ancestral relationship between aberrations $A_{k2}$ and $A_{k3}$. In the first solution $A_{k2}$ and $A_{k3}$ must be on separate branches, in the second solution $A_{k2}$ must be an ancestor of $A_{k3}$, and in the third solution $A_{k3}$ must be an ancestor of $A_{k2}$. 
\end{description}
The number of solutions grows dramatically with $S_k$ ($16$ solutions for $S_k=4$, $125$ solutions for $S_k=5$, $1296$ solutions for $S_k=6$). Therefore, herein we explicitly show the solutions for $S_k<4$ and implement the method to adress practical scenarios, such as nucleotide point mutations ($S_k\le 4$).

\begin{figure}[!ht]
\begin{center}
\includegraphics[width=6in]{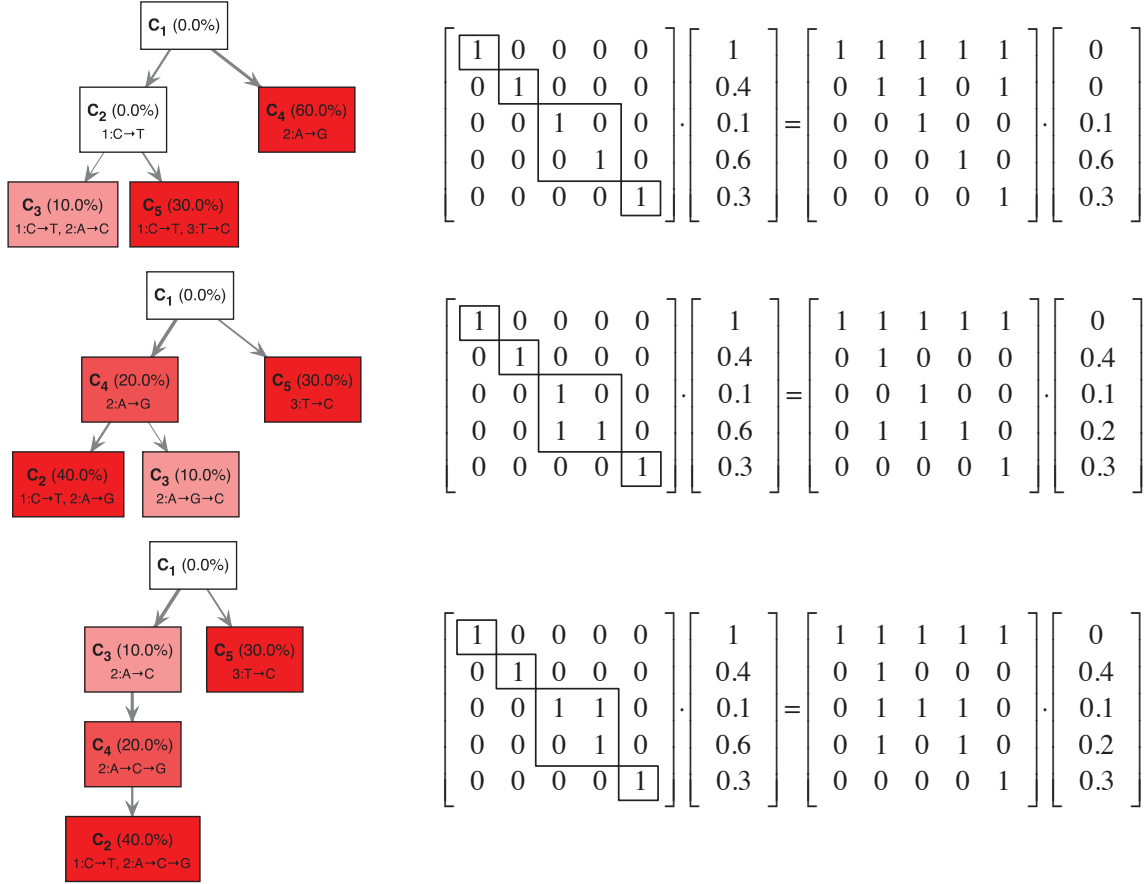}
\end{center}
\caption{
{\bf Three solutions of the generalized subclonal deconvolution problem for a mixture of three sequences in presence of poly-allelic mutations.} We analyzed an aggregate sample composed of three subclones with sequences "TCT", "TAC" and "CGT" mixed with frequencies $0.1$, $0.3$ and $0.6$, respectively. In this example, the wildtype sequence  "CAT" is absent in the mixture. Nonzero elements of the aggregate frequency matrix $\mathbf{Z}$ are concatenated in the $\mathbf{z}$ vector, which consists of the elements $z_{1C}=0.6$, $z_{1T}=0.4$, $z_{2A}=0.3$, $z_{2C}=0.1$, $z_{2G}=0.6$, $z_{3C}=0.3$ and $z_{3T}=0.7$. There are three binarization matrices ($\mathbf{B^{(1,1,1)}}$, $\mathbf{B^{(1,2,1)}}$ and $\mathbf{B^{(1,3,1)}}$) to Equation~\eqref{EqB} and one solution for each binarization matrix is shown. Mutations are shown using the notation $"\text{position}\!:\!\text{reference}\!\rightarrow\!\text{mutated}"$, e.g. the notation $2\!:\!A\!\rightarrow\!G$ indicates that nucleotide at position $2$ was mutated from Adenine to Guanine. The notation $2\!:\!A\!\rightarrow\!G\!\rightarrow\!C$ indicates that nucleotide at position $2$ was first mutated from Adenine to Guanine and then from Guanine to Cytosine. Top: solution based on the binarization matrix $\mathbf{B^{(1,1,1)}}$, in which the subclones $C_3$ and $C_4$ associated with the aberration events $A_{2C}$ and $A_{2G}$ are on separate branches; Middle: solution based on the binarization matrix $\mathbf{B^{(1,2,1)}}$, in which the aberration event $A_{2G}$ (subclone $C_4$) happens before of $A_{2C}$ (subclone $C_3$; bottom: solution based on the binarization matrix $\mathbf{B^{(1,3,1)}}$, in which the aberration event $A_{2C}$ (subclone $C_3$) happens before $A_{2G}$ (clone $C_4$). The solutions are shown both as evolutionary trees (left) and in matrix form according to Equation~\eqref{EqGeneralModel}.  }
\label{FigCATSol}
\end{figure}

The set of all possible solutions to Equation~\eqref{EqB} is given by the Cartesian product of the solution sets of each subproblem. This implies that the number of solutions of Equation~\eqref{EqB} is $\prod_{k=1}^K S_{k}^{S_{k}-2}$. We denote by  $\mathbf{B^{(t_1,\ldots,t_K)}}$ the block matrix solution associated with the blocks $\mathbf{B_1^{t_1}}, \dots, \mathbf{B_K^{t_K}}$ representing the solutions of each subproblem. Similarly, we denote by $\mathbf{y^{(t_1,\ldots,t_K)}}$ the aggregate signal vector given by $\mathbf{y^{(t_1,\ldots,t_K)}} = \mathbf{B^{(t_1,\ldots,t_K)}} \mathbf{z}$. It is possible to reduce the number of rows and columns in Equation~\eqref{EqGeneralModel} by substituting all the $K$ rows corresponding to the unmutated states (shown in green in Figure~\ref{FigCAT}) with a single dummy wildtype aberration (shown in blue in Figure~\ref{FigCAT}). In our toy example, one aberrated state is observed at positions $1$ and $3$ ($S_1=S_3=2$), while two aberrated states ($C$ and $G$) are observed at position $2$ ($S_2=3$). Therefore, there are three solutions to Equation~\eqref{EqB}: $\mathbf{B^{(1,1,1)}}$, $\mathbf{B^{(1,2,1)}}$ and $\mathbf{B^{(1,3,1)}}$. Figure~\ref{FigCATSol} illustrates the best solution for each of these binarization matrices in a graphical and matrix form. The solution associated with $\mathbf{B^{(1,1,1)}}$ is the only TrAp-solution of the generalized subclonal deconvolution problem since it has the minimum number of populated subclones (sparsity constraint).

In summary the generalized subclonal deconvolution problem can be solved as follows:
\begin{enumerate}
\item Vectorize the aggregate frequency matrix $\mathbf{Z}$ and identify all binarization matrices $\mathbf{B^{(t_1,\ldots,t_K)}}$ (Equation~\eqref{EqB}) compatible with the vector $\mathbf{z}$.
\item For each binarization matrix $\mathbf{B^{(t_1,\ldots,t_K)}}$, identify all first generation trees from the aggregate signal vector $\mathbf{y^{(t_1,\ldots,t_K)}} = \mathbf{B^{(t_1,\ldots,t_K)}}  \mathbf{z}$ and combine the first generation trees to generate all partial trees compatible with $\mathbf{B^{(t_1,\ldots,t_K)}}$.
\item Discard partial trees that do not have the minimum number of populated subclones. 
\item Generate all evolutionary trees consistent with the partial trees comprising a maximum number of first generation trees. This step is performed as described above, but with the additional constraint that $c_{ks1,ks2}=b_{ks1,ks2}$ for any pair of states $s_1$ and $s_2$ at position $k$, where $1\le k \le K$, $1 \le s_1 \le S_k$ and $1 \le s_2 \le S_k$.

\item Optimize the shallowness constraint by sorting the generated solutions by the depth of the generated tree.
\end{enumerate}

\subsection*{Benchmarks using simulated data}
The models presented above show that TrAp is an efficient algorithm for inferring subclonal components from the aggregate measure. In particular we have shown that in the absence of noise TrAp returns the exact solution when the underlying subclonal population satisfies reasonable constraints and that the algorithm is always able to find at least one solution. However, experimental measurements are often noisy and can only have finite precision. In this section, we first discuss two approaches to treat noisy input and then we apply our algorithm to simulated aggregate signals from random {\it in silico} trees showing that TrAp is robust to typical noise levels found in genomic experiments.

\subsubsection*{Handling Measurement Errors}
In order to accommodate different types of noise that may arise in genomic data, we incorporated two error models in the TrAp algorithm. In both error models, the input to the TrAp algorithm requires an additional vector $\mathbf{\varepsilon}$ of size $N$ whose elements $\varepsilon_i$ are related to the precision of the measure $y_i$.
The error related to the dummy variable is denoted by $\varepsilon_1$ and is set to $0$ as $y_1=1$ is a constraint of the model and thus $\varepsilon_1$ must vanish.

First, we examine the \textbf{bound error model} in which we assign a threshold to the error $\varepsilon_i \ge 0$ of every underlying aggregate signal $\tilde{y_i}$ such that each measured signal ${y_{i}}$ will be in the range $\left[\tilde{y_{i}}-\varepsilon_{i},\tilde{y_{i}} + \varepsilon_{i}\right]$. Equation~\eqref{Eqxiphi0} is then modified accordingly and we can state that the subclone $C_i$ defined by aberration $i$ is not populated if and only if:

\begin{equation}\label{Eqxierr}
  \left| y_i - \sum_{j=1}^{N} \phi_{i j} y_j  \right| \le \varepsilon_i + \sum_{j=1}^{N} \phi_{i j} \varepsilon_j. 
\end{equation}

Next, we implement a \textbf{normal error model} assuming normally distributed measurements errors using a confidence interval approach to determine whether a subclone is populated or not.  Specifically, we assume that the underlying aggregate signal is normally distributed around the observed signal, i.e. $y_i  \sim \mathcal{N}\left(\tilde{y_i} ,\varepsilon_i^2 \right)$. We substitute each term of the left-hand side of Equation~\eqref{Eqxiphi0} by its normal distribution in order to derive the distribution $r \sim \mathcal{N}\left(\mu_r , \sigma^2_r\right)$  with mean $\mu_r = \tilde{y_i} - \sum_{j=1}^{N} \phi_{i j} \tilde{y_j}$ and variance $\sigma_r^2 = \varepsilon_i^2 + \sum_{j=1}^{N} \phi_{i j} \varepsilon_j^2$. Using the distribution of $r$ and a desired confidence level $\alpha > 0$ (default $0.05$), we can define that clone $C_i$ is not populated if $0$ falls within the confidence interval $[q_{\frac{\alpha}{2}},q_{\frac{1-\alpha}{2}}]$, where $q_{\alpha}$ is the $\alpha$ quantile of the distribution of $r$. 

Once the error model is chosen, the algorithm generates all optimal TrAp-trees in a similar fashion to the noise-free case. The main difference is that in the first step of the TrAp algorithm, Equation~\eqref{Eqxierr} (or a confidence interval on the distribution $r \sim \mathcal{N}\left(\mu_r , \sigma^2_r\right)$) is used instead of Equation~\eqref{Eqxiphi0} to identify first generation trees. Moreover, instead of using back-substitution for finding the vector $\mathbf{x}$, we solve the nonnegative linear least square problem $\mathbf{C} \mathbf{x}=\mathbf{y}$ with the constraint $x_k=0$ for all non-populated subclones $k$ associated with the parents of the first generation trees. This fitting allows us to obtain a value of exactly zero for all non-populated subclones and to distribute measurement error more evenly in the vector $\mathbf{x}$.

\subsubsection*{Deconvolution of simulated noisy aggregates}
To confirm that TrAp can correctly infer the subclonal composition from aggregate noisy signals with typical noise levels found in genomic experiments, we performed simulations starting with random {\it in silico} evolutionary trees (see Methods) with different numbers of aberrations $N$ and different numbers of populated subclones $P$. For each tree, we also studied the effect of different magnitudes of measurement errors $\varepsilon$ and we investigated the operating conditions for which TrAp would assign the best score to the true solution. 

For each aggregate signal from a random tree we examined: (i) whether the true solution (i.e. the solution associated with the simulated tree), which is by construction one of the possible solutions to the subclonal deconvolution problem, had the minimum number of populated subclones among all solutions (sparsity constraint), (ii) whether the true solution had the minimum number of populated subclones and minimum number of levels of the evolutionary tree among all solutions generated by TrAp (sparsity and shallowness constraints), and (iii) whether the true solution was the only TrAp solution (sparsity and shallowness constraints and uniqueness of the optimal solution). 

\begin{figure}[!ht]
\begin{center}
\includegraphics[width=4in]{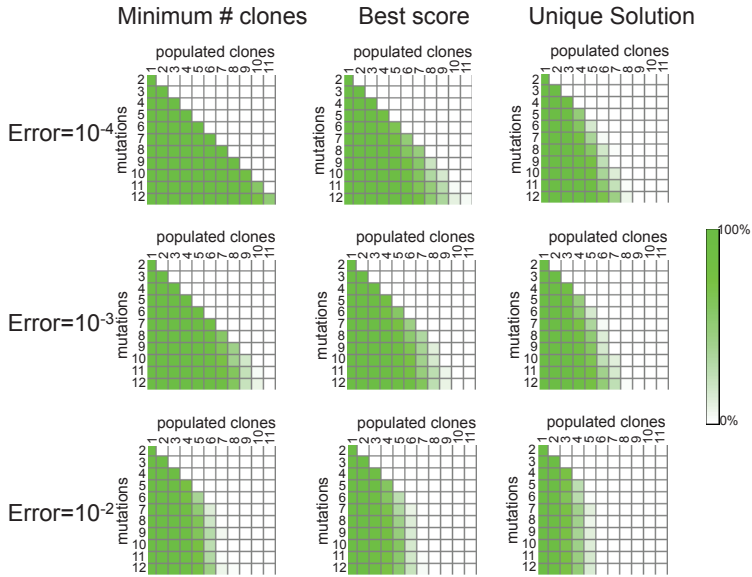}
\end{center}
\caption{
{\bf Deconvolution of simulated data.} 
We generated $1000$ simulations (see Methods) for each combination of number of populated subclones (columns), number of mutations (rows). We performed this analysis using different level of noise (error)  drawn from a uniform distribution $\mathcal{U}\left(-\varepsilon,\varepsilon\right)$. The heatmaps (tables) show the percentage of trees in each cell  in which the TrAp solution has the minimum number of subclones (left panel), is also the shallowest solution with the best score (middle panel) and in addition to these two conditions is also unique (right panel).
If the best solution is unique.}
\label{FigGreen}
\end{figure}

The results of the simulations show that aggregate signals from sparse trees are deconvolved correctly even in presence of typical noise levels of sequencing experiments (Figure~\ref{FigGreen}). We note that for simulations of non-sparse trees, TrAp generates a large number of possible solutions of which only one is the true solution. Furthermore, in the presence of high levels of noise, the TrAp algorithm identifies a large number of first generation trees that satisfy Equation~\eqref{Eqxierr} and generates solution trees whose number of populated subclones is smaller than the number of populated subclones of the true solution.

\subsection*{Analysis of simulated mixtures of biological data}

\subsubsection*{Deconvolution of a mathematical mixtures of karyotyping data from single tumor biopsies}

\begin{figure}[!ht]
\begin{center}
\includegraphics[width=4.8in]{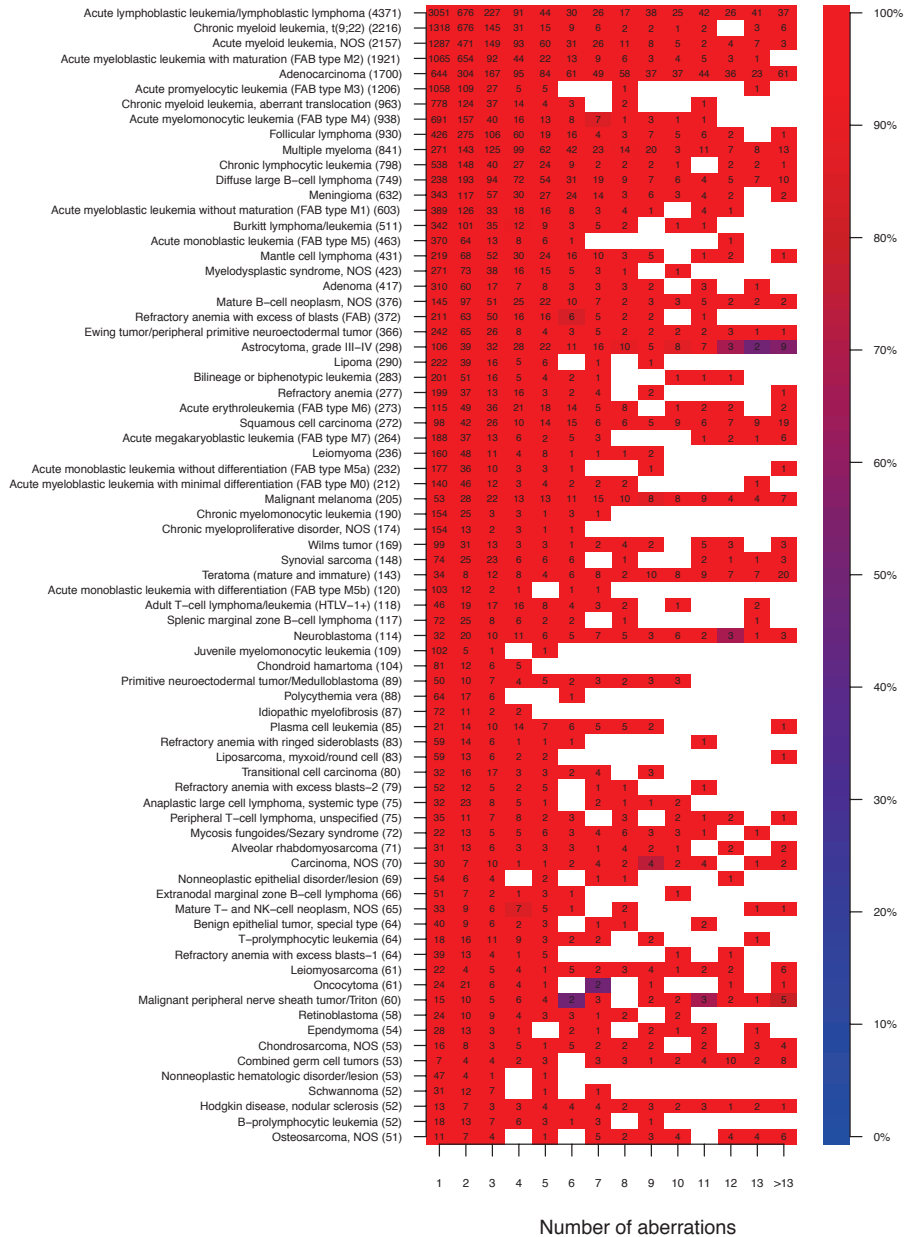}
\end{center}
\caption{
{\bf Applicability of the TrAp algorithm for different number of aberrations events and different kind of tumors.}  
Each entry in the heat map shows the number of biopsies, which contained a given number of aberrations. Each cell is colored by the fraction of biopsies for which TrAp could reconstruct the correct composition of the subclones, from red (all biopsies could be reconstructed) to blue (no biopsies could be reconstructed). The constraints required by the TrAp algorithm are satisfied in most cancer types, with the exception of astrocytoma  of grades III and IV.
}
\label{FigMitelmanApplicability}
\end{figure}

After showing that our approach can correctly deconvolve aggregate signals of subclones with a tree-like genealogy, we sought to investigate whether actual subclonal populations can be charted on evolutionary trees. For this purpose we analyzed the Mitelman database, consisting of cytogenetic analyses of more than $60,\!000$ biopsies (see Methods). For each tumor type we counted how frequently the relationships between cancer clones from the same biopsy could be explained by an evolutionary tree that follows the evolutionarity and parsimony constraints (but not necessarily the sparsity and shallowness constraints). We found that almost all biopsies in the Mitelman database can be represented by evolutionary trees (Table~\ref{mitelmanSols}), with the notable exception of astrocytoma of grades III and IV (Figure~\ref{FigMitelmanApplicability}). 

\begin{table}[ht]
\caption{\bf{Applicability of the TrAP algorithm for different number of aberrations events and underlying subclones.}}
\begin{tabular}{|c|c|c|c|c|c|}
	&1 &	2 &	3 &	4 &	$>$4 \\
\hline
1	&100\% (19078)	&n/a	&n/a	&n/a	&n/a\\
2	&100\% (5150)	&100\% (923)	&\textbf{0\% (3)}	&n/a	&n/a\\
3	&100\% (1830)	&100\% (367)	&94\% (83)	&\textbf{0\% (2)}	&n/a\\
4	&100\% (991)	&100\% (182)	&89\% (27)	&89\% (18)	&n/a\\
5	&100\% (656)	&100\% (120)	&88\% (33)	&100\% (8)	&100\% (5)\\
6	&100\% (445)	&100\% (66)	&92\% (13)	&100\% (6)	&50\% (4)\\
7	&100\% (333)	&100\% (58)	&89\% (9)	&25\% (4)	&100\% (2)\\
8	&100\% (241)	&100\% (37)	&86\% (7)	&100\% (3)	&50\% (2)\\
9	&100\% (228)	&100\% (26)	&60\% (5)	&0\% (1)	&100\% (1)\\
10	&100\% (174)	&100\% (14)	&100\% (2)	&n/a	&50\% (2)\\
11	&100\% (196)	&100\% (25)	&67\% (3)	&67\% (3)	&n/a\\
12	&100\% (156)	&100\% (16)	&100\% (3)	&0\% (1)	&50\% (2)\\
13	&100\% (137)	&100\% (21)	&50\% (2)	&n/a	&100\% (1)\\
14	&100\% (94)	&100\% (12)	&n/a	&100\% (1)	&100\% (1)\\
$>$ 14	&100\% (152)	&100\% (22)	&57\% (7)	&25\% (4)	&25\% (4)\\

%table information
\end{tabular}
\begin{flushleft} We mathematically mix all karyotypes of each single patient from the Mitelman database and apply the TrAp algorithm for each of these mixtures. The ability of the TrAp algorithm to extract the correct underlying clonal or subclonal components depends on the number of actual components (columns) and the multiplicity of aberrations studied in each mixture (rows). The frequency in which TrAp is able to recover the correct underlying components is shown in percentages. The number of mixtures for a given size of aberration multiplex (row) and given number of actual underlying components (column) is shown in parentheses. Note that when the column index is greater than the row index (entries shown in bold) the parsimony constraint cannot be satisfied. 
\end{flushleft}
\label{mitelmanSols}
\end{table}

\begin{figure}[ht]
\begin{center}
\includegraphics[width=6in]{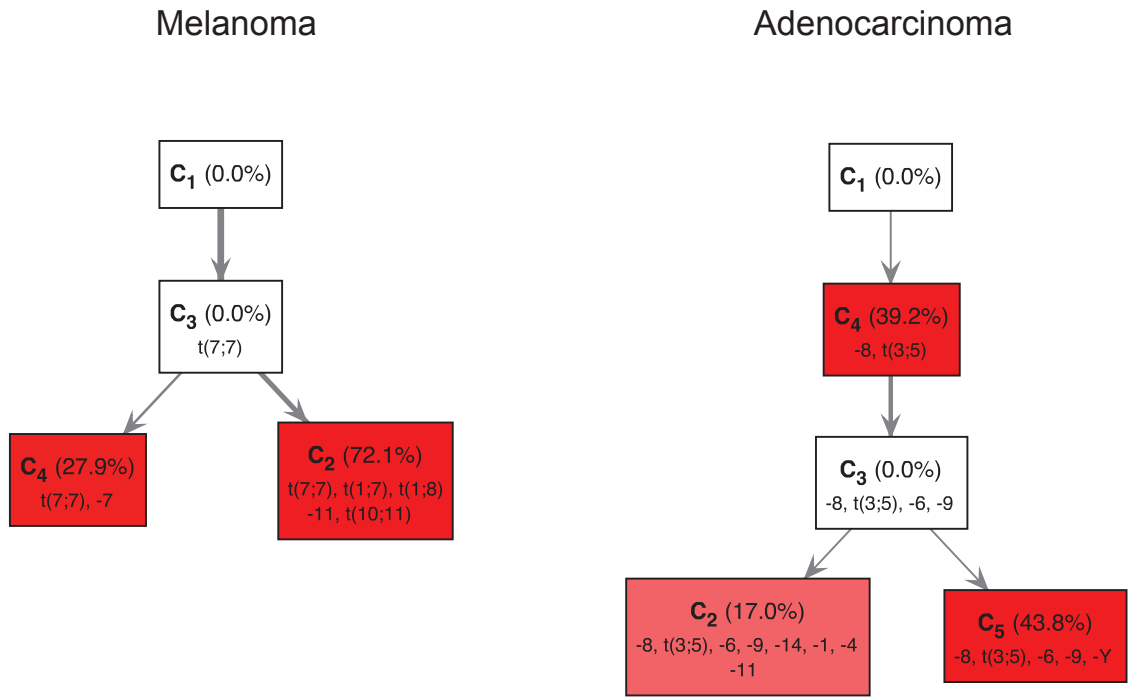}
\end{center}
\caption{
{\bf Deconvolution of random mixtures of three subclones.}  
The boxes represent different subclones, each denoted by the list of its aberrations. The aberration profiles of two subclones identified by cytogenetics in a melanoma biopsy (left) and the aberration profiles of three subclones identified in an adenocarcinoma biopsy (right) have been mixed \textit{in silico} using random coefficients. In both case, the mixtures were successfully deconvolved. Aberrations are grouped within the boxes according to the order of occurrence. The reconstructed evolutionary trees suggest intermediate (white boxes), probably rare, subclones that were not reported in the cytogenetic data.
}
\label{FigRed}
\end{figure}

Next, we investigated whether the TrAp algorithm could uniquely deconvolve mixtures of the cancer subclones within a biopsy. As these aggregate signals are obtained by mixing actual subclonal profiles, we consider these signals to be more realistic than our previous {\it in silico} simulations. For each biopsy, we generated  {\it in silico} mixtures by combining the cytogenetic profiles of each subclone using random nonnegative coefficients (see Methods). We then applied our TrAp method to deconvolve \textit{in silico} mixtures of these cancer clones and found that TrAp could correctly deconvolve $99.7\%$ of these realistic aggregate signals. However, we note that this task was trivial for a significant fraction of the biopsies whose number of aberrations and/or subclones is small. Moreover, the TrAp algorithm inferred also intermediate nodes in the evolutionary tree that did not correspond to any of the cytogenetic profiles for the biopsy, providing a plausible picture of the evolutionary order in which the aberrations occurred. Figure~\ref{FigRed} shows the result of two deconvolution simulations, one from a melanoma sample with $2$ subclonal populations \cite{Melanoma:figred}  and one from an adenocarcinoma sample with $3$ subclonal populations \cite{adenocarcinoma:figred}. 
An interesting, yet rare (only $5$ examples, shown in bold in Table~\ref{mitelmanSols}), case of biopsies showed more clones than aberrations. Albeit a tree-like genealogical relationship can be constructed, these biopsies do not satisfy the parsimony constraint since the number of subclones $M$ is greater than the number of mutations $N$. For this reasons, their genealogy cannot be reconstructed by the TrAp algorithm or by any other method that makes use of a similar parsimony constraint \cite{farris1966estimation,kluge1969quantitative,MaxPars,clement2000tcs}. 

\subsubsection*{Deconvolution of a mathematical mixture of somatic hypermutation data with poly-allelic mutations in a single nucleotide}

Somatic hypermutation (SHM) introduces mutations that target the variable regions associated with immune adaptivity, i.e. \textit{Ig} loci. In particular, SHM involves a programmed process of mutations that affects the variable regions of immunoglobulin genes and starts from an initial dividing single cell (a na\"{i}ve B cell in this case). All descendants of the founder cell accumulate mutations and, at the same time, are subjected to a strong selective pressure. For this reason, SHM is a particularly good biological model system to test our deconvolution method, which imposes tree-like evolutionary constraints.

We considered a dataset where a total of $20$ mutated nucleotides in the V(D)J region of the \textit{Ig} locus were measured in $8$ sequences extracted from the same germinal center (see Methods) \cite{KleinisteinGC}. This dataset was particularly interesting because of the high number of mutations found and because of the presence of poly-allelic mutations. 

\begin{figure}[!ht]
\begin{center}
\includegraphics[width=\linewidth]{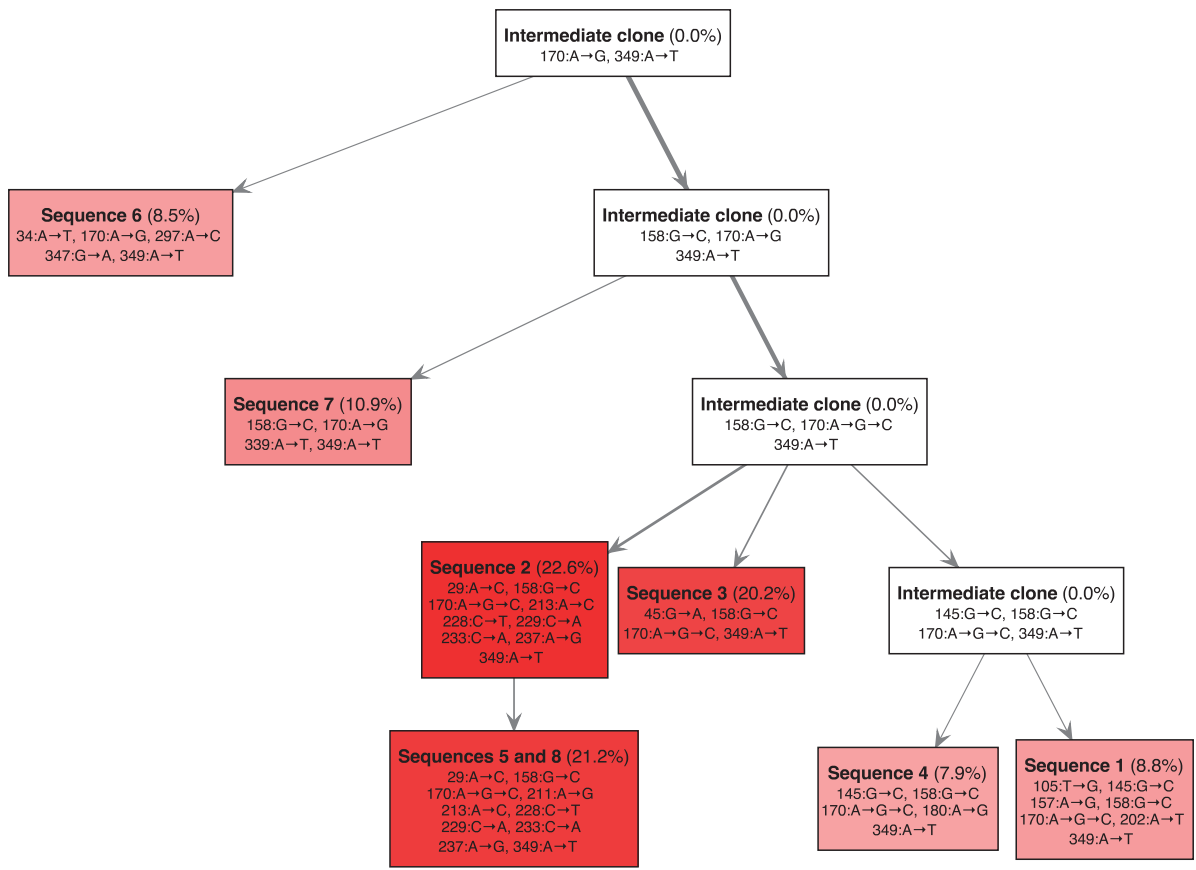}
\end{center}
\caption{
{\bf Deconvolution of a random mixture of eight sequences from somatic hypermutation data.} Eight sequences from the \textit{Ig} locus of eight cells extracted from the same germinal center were mixed with the random coefficients given by $\mathbf{x}=\left[ 8.8\%, 22.6\%, 20.2\%, 7.9\%, 5.7\%, 8.5\%, 10.9\%, 15.5\%\right]$. Since sequences $5$ and $8$ are identical, they are grouped in a single clone whose relative frequency is $5.7\% + 15.5\% = 21.2\%$. In total, twenty mutated nucleotides were found in the data and two different mutations were identified at position $170$. Mutations are shown using the notation $"\text{position}\!:\!\text{reference}\!\rightarrow\!\text{mutated}"$, e.g. the notation $170\!:\!A\!\rightarrow\!G$ indicates that the nucleotide at position $170$ was mutated from Adenine to Guanine. The notation $170\!:\!A\!\rightarrow\!G\!\rightarrow\!C$ indicates that the nucleotide at position $170$ was mutated twice, first from Adenine to Guanine and then from Guanine to Cytosine. In this example, all seven subclones were correctly deconvolved by the TrAp algorithm, the frequency of the subclones was correctly estimated and the solution was unique. 
}\label{FigSHM}
\end{figure}

We mathematically mixed the multi-subclonal data and applied our TrAp algorithm taking into account that the SHM scenario  consists of non-binary aberrations. In all simulations, TrAp was able to recover the original sequences and the solution was unique in $65\%$ of the simulations. The TrAp-solution of one simulation is shown in Figure~\ref{FigSHM}. However, even if the solution was not always unique, in more than $97\%$ of the simulations there were only five or less candidate solutions satisfying the evolutionary, parsimony and sparsity constraints, all of which correctly identified six out of seven subclones. 

In addition to the identification of the underlying subclones, the TrAp algorithm generates evolutionary trees, which represent the B cell lineage during the somatic hypermutation process. The reconstruction of B cell lineage can provides important insights into the mechanisms that regulate adaptive immunity.  B cell lineage reconstruction is generally performed using maximum parsimony constraints \cite{clement2000tcs} using the sequences of several \textit{Ig} loci as input. However, in contrast to these approaches, the TrAp algorithm is able to generate maximum parsimony trees when only the relative frequency of mutations at each nucleotide is available. Therefore, the TrAp algorithm can be used to generate parsimonious evolutionary trees when only partial sequence information is available, e.g. when only short read sequences from a single aggregate sample are available, or when the loci analyzed span a region that is too large to be fully sequenced. 

\subsection*{Analysis of tumor biopsies}

\subsubsection*{Comparison between subclonal aberration profiles inferred from heterogeneous cell populations and singl-cell aberration profiles}

We analyzed data from a recent study on renal cell carcinoma where two aggregate samples and twenty single cells were isolated from a clear cell renal cell carcinoma (ccRCC) and subjected to exome sequencing. Interestingly, the original study only showed partial similarity between the single cells and the aggregate \cite{SingleCell25}. However, since the single cells and the aggregate used in the experiments are from the same tumor, we sought to investigate whether any subclones inferred by TrAp would share a similar combination of mutations found in any of the single cells.

We applied our TrAp algorithm to the aggregate sample and obtained an evolutionary tree consisting of three main subclones. Due to the lack of extensive validations, we limited ourselves to investigate whether mutations that co-occur in the TrAp-solution also co-occur in single cell samples. We considered the mutations that were validated by bioinformatics analysis (Table S3A from Xu \textit{et al.} \cite{SingleCell25}) and by PCR validation (TableS3B). The fraction of correctly estimated co-occurrences was $0.76$ for mutations validated by bioinformatics analysis and $0.74$ for mutations validated by PCR. 

\subsubsection*{Melanoma data}

Finally, we applied our algorithm to investigate evolutionary mechanisms in tumor metastases using exome sequencing data from three tumor metastases (TM1: left lateral chest wall, TM4: pleural cavity, and TM3: right axilla) a matched normal sample (N: left lateral chest wall) of one melanoma patient \cite{Spore:Exome}. TrAp can efficiently handle aggregate signal vectors of about $20$ unique frequencies and therefore we perform deconvolution analysis only on one chromosome. We selected chromosome $18$ as it contains the tumor suppressor Deleted in Colorectal Cancer (\textit{DCC}) gene, which is known to exhibit a high load of mutations only in melanoma \cite{DCC}, in contrast to low expression, loss of heterozygosity or epigenetic silencing in other tumors. 

To apply the TrAp algorithm, we first preprocessed the data and selected $19$ mutations on chromosome $18$ (see Methods). We labeled each mutation according to the gene affected and the amino acid change caused by the mutation (e.g. the label DCC.L1099H indicates a mutation in the \textit{DCC} gene that causes a mutation from a Leucine to a Histidine at position $1099$ in the DCC protein). There are six mutations with $>99\%$ frequency in all samples (including the normal): ADNP2.G54G, ALPK2.I2157V, CD226.S307G, DCC.F23L, NETO1.S481N, and SLC39A6.E119D. The only other mutation found in the normal sample was TCEB3CL.S301C, which occurs with frequency $>90\%$ in all samples. Moreover, the mutations ALPK2.R136S, CHST9.S122N, FAM38B.V2463, LAMA1.S1577A, LAMA1.K2002E, MYOM1.T215M, SERPINB10.R246C and {SLC14A2.A880T} were found in all three tumor samples and shared a similar frequency profile. The mutations DSC3.A28D, DSG1.M11V and IMPACT.D125E were found only in the metastases samples TM3 and TM4 and shared a similar frequency profile. Finally, the mutation DCC.L1099H was found only in the sample TM3. 

\begin{figure}[!ht]
\begin{center}
\includegraphics[width=\linewidth]{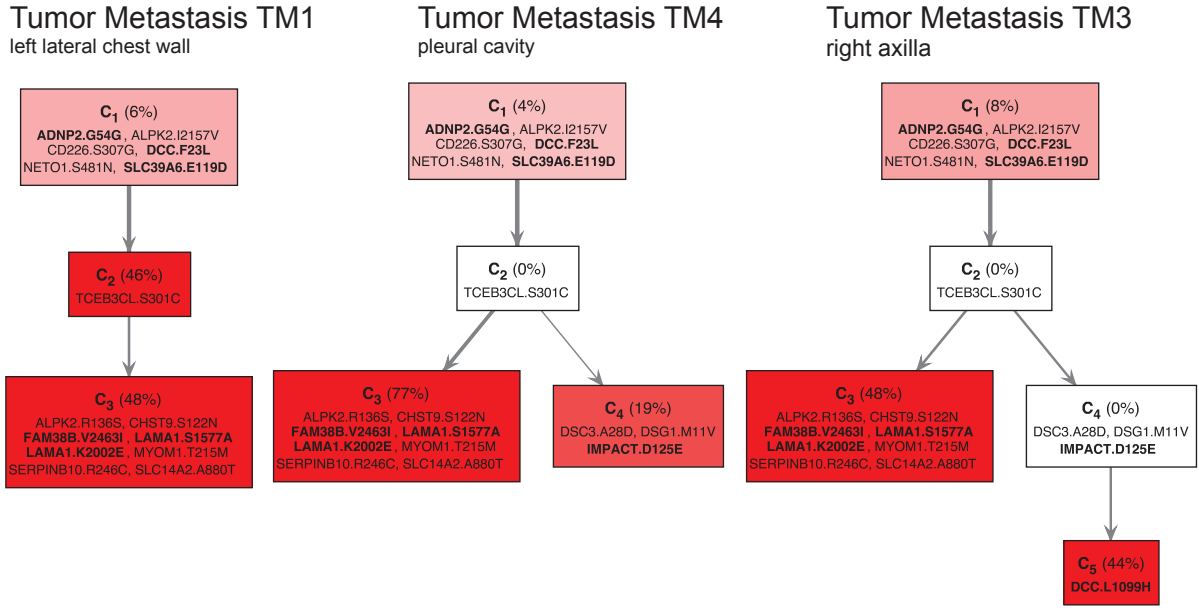}
\end{center}
\caption{
{\bf Evolutionary trees inferred from three metastases of a melanoma patient.} Each subclone in these trees is represented by a box with a list of mutations that includes only its new mutations (ancestral mutations can be read off by tracing back the mutation lists of all of its ancestors). Mutations are labeled according to the gene affected and the amino acid change caused by the mutation (e.g. the label DCC.L1099H indicates a mutation in the \textit{DCC} gene that causes a mutation from a Leucine to a Histidine at position $1099$ in the DCC protein). Highly expressed genes from this patient are indicated in bold. Mutations in the left branch of TM4 are more abundant than in TM1 and TM3. $44\%$ ($19\%$) of the subclones of TM3 (TM4) have mutations in DSC3, DSG1 and IMPACT. The TM3 subclone has an additional mutation in DCC.}
\label{FigYuhuy}
\end{figure}

Independent runs on the three metastatic samples gave $33$ optimal solutions for TM1, $222$ for TM3 and only $1$ TrAp-solution for TM4. These high number of solutions is due to the substantial noise of the experiment (in the range $0.005 - 0.025$) and the fact that in samples TM1 and TM3, two of the subclones have similar frequencies and are thus difficult to separate from one another. However, TrAp identified an unique solution in the sample TM4, where the three populated subclones are distributed with significantly different frequencies (Figure~\ref{FigYuhuy} middle). Next, we reasoned that the metastatic TM1, TM3 and TM4 samples may share common ancestors and that their evolutionary profiles may be related. We then applied our TrAp algorithm while also imposing that all evolutionary trees must be a subset of one global evolutionary tree. This approach returned an unique solution for each sample, all of which were among the solution sets identified in the previous analyses. We observe that this approach can be very powerful since, in principle, it allows the reconstruction of very large trees by combining several snapshots of the related subclonal populations.

The results of the deconvolution are shown in Figure~\ref{FigYuhuy}. We observe the presence of two main branches. Mutations in the left branch of TM4 (77\%) are more abundant than in TM1 and TM3 (48\%). We note
that LAMA1, a protein that is involved in cell adhesion, is present in the right branch. 44\% (19\%) of the subclones of TM3 (TM4) have mutations in DSC3, DSG1 and IMPACT. The TM3 subclone also acquires a second mutation in the \textit{DCC} gene (DCC.L1099H-L) in addition to the mutation DCC.F23L, which was hereditary.  
The novel mutation in the \textit{DCC} gene occurs close to the boundary between the extracellular domain and the transmembrane domain of the protein product. The resulting Histidine amino-acid is positively charged, opposed to the Leucine amino-acid of wildtype, which is neutrally charged. Because this change is next to the cell membrane, it may have repercussions on the functionality of the DCC protein product, perhaps causing inactivation, similar to the inactivation caused by loss of heterozygosity and transcript suppression observed in other cancer types.

\section*{Discussion}
In the present study we presented the TrAp algorithm, a tool for inferring subclonal composition and abundance from a single aggregate measurement experiment. As we have shown, TrAp is robust to noise and it can deconvolve mixtures where multiple mutations occurs at the same locus. TrAp efficiently enumerates all possible solutions that are compatible with the model constraints, thus always identifying the sparsest and most parsimonious solution(s).  However, TrAp will also generate trees (cf. Equation~\ref{EqCascade}) in cases where no tree structure can be inferred. As we have shown, such structures, while deviating from the true underlying population structure, can still capture relevant co-occurrences of mutations that are specific to certain subclones. Further, in contrast to parsimonious neighbor-joining approaches, which rely on sampling single subclones from the population (e.g. single cell experiments), TrAp utilizes one single aggregate experiment as input, thus overcoming the issue of small sampling size, which may be insufficient to cover the whole spectrum of subclones in a sample. We successfully deconvolved systems of up to $25$ aberrations. Although this number may not always be large enough to consider all somatic mutations found in a tumor sample, this problem can be circumvented by clustering aberrations with similar frequencies before running the TrAp algorithm. 

The level of characterization achieved by subclonal deconvolution holds high potential for personalized therapies. Possible applications include the classification of subclones in primary tumors, the identification of the seeds of metastases, tracing of resistant subclones especially after drug treatments and developing treatment strategies to eliminate resistant subclones. Furthermore, our proposed model can be applied to other medical problems, such as tracing bacterial or viral paths of adaptation within the infected host, detailed genome-wide reconstruction of the epigenetic differentiation program, or class specification in the hematopoietic system or in other systems.

\section*{Methods}

In this section, we describe the procedures used to preprocess input data for the TrAp algorithm.

\subsection*{Data processing and generation}

\subsubsection*{Random \textit{in silico} evolutionary trees}

We performed simulations by sampling genotypes whose size $N$ ranged from $1$ to $12$ and with underlying number of populated subclones $P$ ranging from $1$ to $N-1$. The simulations were repeated for measurement error values $\varepsilon$ equal to $10^{-2}$, $10^{-3}$ and $10^{-4}$. For each combination of these quantities, we performed $1000$ runs using \textit{in silico} generated data as follows: during each run, a random evolutionary tree with $N$ aberrations was generated. The set of $P$ populated subclones was then selected by first including all leaves of the generated tree and then adding the remaining subclones randomly. The frequency of the populated subclones was randomly assigned and the frequency of the non-populated clones was set to zero. Next, the aggregate frequency vector $\mathbf{\tilde{y}}$ was calculated from the generated tree. Finally, we perturbed each element of $\tilde{y_i}$ by adding an error $\varepsilon_i$ drawn from a uniform distribution $\mathcal{U}\left(-\varepsilon,\varepsilon\right)$. The elements  $y_i=\tilde{y_i}+ \varepsilon_i$ and the error $\varepsilon_i$ are used as input for the bound error model option of the TrAp algorithm.

\subsubsection*{Cytogenetic data}

The cytogenetic data was obtained from the Mitelman database, which contains $61,\!846$ biopsies as of August 15, 2012. We accessed the database through the CGAP Website\cite{Mitelman} and we filtered out $29,\!842$ biopsies with uncertain calls (indicated by a "?" in the karyotype data). We focused our search only on aberrations that are binary by nature, namely chromosome deletions and translocations. For each biopsy, we performed $100$ \textit{in silico} simulations in which we mixed the subclones using random nonnegative coefficients.

\subsubsection*{Somatic hypermutation data}
SHM sequencing data was derived from B cells undergoing somatic hypermutation (SHM), a process that leads to high-affinity antibody molecules \cite{SHM1}. In details, we analyzed sequences of the V(D)J region extracted from the same germinal center from the sample \textit{11930d16\_4print.2}, which was sequenced by Anderson \textit{et al.} \cite{KleinisteinGC, SHM:laser}. The sequences were aligned using the IMGT alignment tool \cite{IMGT, lefranc2005imgt} in order to properly align the V, D and J regions. We selected $8$ sequences that were aligned to the same V(D)J sequence ($V_1(D_1)J_1$). Since these sequences are from the same germinal center and align to the same V(D)J sequence, they are expected to stem from a single na\"{i}ve B cell and have evolved through the somatic hypermutation process. From the sequencing experiment, $20$ mutated nucleotides were identified. Furthermore, a polyallelic mutations were found at position $170$, where both $A\!\rightarrow\!C$ and $A\!\rightarrow\!G$ mutations were observed. Next, we considered the $7$ unique sequences (sequences $5$ and $8$ were identical) as representatives of the genome of $7$ different subclones. Similarly to the previous application, we mixed these subclones using random nonnegative coefficients and performed $1000$ simulations using random nonnegative coefficients.

\subsubsection*{Exome capture sequencing data}
Exome-capture data \cite{Exome1} was obtained from a recent clear cell renal cell carcinoma (ccRCC) study \cite{SingleCell25} and from the melanoma patient YUHUY of the Yale cohort, for which DNA from normal circulating lymphocytes and three metastases (TM1, TM3 and TM4) were subjected to exome-capture Illumina Hi-Seq sequencing \cite{Spore:Exome}.

Exome-Seq reads from the aggregate samples of the ccRCC patient were combined and aligned to the human reference genome (assembly hg19) using Bowtie \cite{bowtie} with parameters "-n3  -k1 -m20 -l32 -{}-chunkmbs 1024  -{}-best -{}-strata". The frequency and coverage of each point mutation was computed using VarScan \cite{VarScan}. We further selected the mutations that were validated by Xu \textit{et al.} \cite{SingleCell25} by PCR validation (TableS3B) and by bioinformatics analysis (Table S3A ), whose genomic coordinated were realigned to the assembly hg19 using the Lift-Over tool of Galaxy \cite{Galaxy1}. 

For the melanoma patient YUHUY \cite{Spore:Exome}, we selected $19$ mutations that were homozygous in the normal sample (i.e. $y\approx 0$ or $y \approx 1$), had high sequence coverage (i.e. more than $200$ reads covering the specific nucleotide) and were localized on chromosome $18$.

Since none of the genomic positions analyzed contained poly-allelic mutations, we assign a binary state (normal/mutated) to each selected genomic position and we estimated the aggregate signal and measurement error for each mutation event using a normal approximation as 
\begin{equation}\label{Eqnorm}
y_i= \frac{m_i}{n_i}; \varepsilon_i = \sqrt{\frac{y_i  (1-y_i)}{n_i}}
\end{equation}
where $n_i$ is total the number of reads covering position $i$ and $m_i$ is the number of reads covering position $i$ where the nucleotide $i$ is mutated. Finally, the $\mathbf{y}$ and $\mathbf{\varepsilon}$ vectors were used as input for the TrAp algorithm, which was set to use the normal error model.

\subsection*{Implementation of the TrAp algorithm}
TrAp was programmed in Java. TrAp makes use of the Java Matrix package \cite{jama} for linear regression and code by Josh Vermaas to solve the nonnegative least square problem using JAMA. The Java Universal Network/Graph Framework (JUNG) \cite{JUNG} is used for creating pictorial representations of evolutionary trees. TrAp is released under the GNU Lesser General Public License 3.0 and can be downloaded on the SourceForge repository of Kluger's lab at the URL \url{http://sourceforge.net/projects/klugerlab/files/TrAp/}.

% Do NOT remove this, even if you are not including acknowledgments
\section*{Acknowledgments}

We are thankful to Ruth Halaban, Michael Krauthammer and Perry Evans for providing the melanoma sequencing data. We are grateful to Steven Kleinstein, Mohamed Uduman and Gur Yaari for providing the somatic hypermutation data and for helpful discussions. Finally, we thank Jiaqi Jin and Kelly Stanton for critical readings of the manuscript.

% Bibliography
\newpage
\bibliography{TrAp}

\end{document}